\pdfoutput=1

\documentclass[12pt]{article}
\usepackage{amsmath,epsf,amssymb,latexsym,cite,shadow,eucal,theorem,enumerate}
\usepackage[matrix,arrow,frame,import,curve,color]{xy}
\usepackage{tikz}
\usetikzlibrary{arrows}
\usetikzlibrary{decorations.pathmorphing}
\usetikzlibrary{decorations.markings}

\usepackage{graphicx}

\newtheorem{prop}{Proposition}

{\theorembodyfont{\rmfamily}}

\newif\iffigs\figstrue

\DeclareFontFamily{U}{rsf}{}
\DeclareFontShape{U}{rsf}{m}{n}{
  <5> <6> rsfs5 <7> <8> <9> rsfs7 <10-> rsfs10}{}
\DeclareMathAlphabet\Scr{U}{rsf}{m}{n}

\def\O{\Scr{O}}
\def\C{{\mathbb C}}
\def\P{{\mathbb P}}

\def\R{{\mathbb R}}
\def\Z{{\mathbb Z}}

\def\Im{\operatorname{Im}}
\def\Hom{\operatorname{Hom}}

\def\End{\operatorname{End}}

\def\coker{\operatorname{coker}}
\def\supp{\operatorname{supp}}

\def\Tr{\operatorname{Tr}}

\def\GU{\operatorname{U{}}}

\def\rank{\operatorname{rank}}

\def\p{\partial}

\def\CY{Calabi--Yau}
\def\LG{Landau--Ginzburg}

\def\cA{{\Scr A}}

\def\cI{{\Scr I}}

\def\cP{{\Scr P}}

\def\cE{{\Scr E}}

\def\cG{{\Scr G}}

\def\DC{\mathbf{D}}

\def\ff#1#2{{\textstyle\frac{#1}{#2}}}

\def\mf#1{\mathfrak{#1}}

\def\poso#1{#1\save="x"!LD+<0pt,-0.5mm>;
  "x"!RD+<0pt,-0.5mm>**\dir{.}\restore}

\def\pz{\phantom{-}}

\begin{document}

\begin{titlepage}
\begin{flushright}
DUKE-CGTP-09-01\\
May 2009
\end{flushright}
\vspace{.5cm}
\begin{center}
\baselineskip=16pt
{\fontfamily{ptm}\selectfont\bfseries\huge
Probing Geometry with Stability Conditions\\[20mm]}
{\bf\large  Paul S.~Aspinwall
 } \\[7mm]

{\small

Center for Geometry and Theoretical Physics, 
  Box 90318 \\ Duke University, 
 Durham, NC 27708-0318 \\ \vspace{6pt}

 }

\end{center}

\begin{center}
{\bf Abstract}
\end{center}
The notion that the geometry of spacetime is given by the moduli space
of 0-branes is examined in four examples of \CY\ threefolds. An
important consideration when determining the moduli space of D-branes
is the stability condition and this is key in our analysis. In the
first two examples, the flop and the orbifold blowup, no surprises are
found. Next, an exoflop is considered where the linear sigma model
implies a $\P^1$ external to the \CY\ threefold is part of the
geometry. The 0-brane probe sees no such external $\P^1$ and
furthermore exhibits a surprising discontinuity when following an
extremal transition associated to the exoflop. Finally we consider a
hybrid model of a Landau--Ginzburg fibration over a $\P^1$. Using the
technology of matrix factorizations we find a D-brane probe whose
moduli space is this $\P^1$ but it is not a 0-brane and is not stable
at the large radius limit of the \CY\ manifold.

\end{titlepage}

\vfil\break


\section{Introduction}    \label{s:intro}

Spacetime is supposed to be ``emergent'' in string theory. The notion
of D-branes in the worldsheet description of string theory is one way
to achieve this idea. If one has a perfect knowledge of the worldsheet
physics then one should understand all possible boundary conditions,
i.e., D-branes in the classical (zero string coupling) limit. Picking
a suitable class of D-brane as a candidate 0-brane one can then try to
construct a moduli space for all D-branes in this class. This
resulting space should be spacetime.

The resulting spacetime is not unique. As is well known, there can be
different choices of the class of a 0-brane and this leads to T-dual
target spaces.

In the case of B-type topological field theories with target space a
compact \CY\ threefold $X$, the set of D-branes is given by the bounded
derived category, $\DC(X)$, of coherent sheaves. Let $\Gamma$ be the
set of D-brane charges, which is known to be given by topological
K-theory \cite{W:K}. The charge associated to a given object in the
derived category is given by the map from the Grothendieck group of
$\DC(X)$ to $\Gamma$:
\begin{equation}
  C:K(X) \to \Gamma.
\end{equation}

Pick a candidate 0-brane charge $\gamma\in\Gamma$. The objects in
$\DC(X)$ corresponding to $C^{-1}(\gamma)$ are far too numerous to
constitute the set of points of any target space $X$. This is
illustrated, for example, by the well-known case of the flop
\cite{Brig:flop}. Suppose $X$ contains a rational curve $C$ which may
be flopped to produce a topologically distinct \CY\ threefold $X'$
containing the flopped curve $C'$. The derived categories $\DC(X)$ and
$\DC(X')$ are equivalent. So $\DC(X)$ contains objects corresponding
to skyscraper sheaves on $C'$ even though $C'$ is not a curve in $X$.

The two-dimensional conformal field theory on the worldsheet does not
allow all the objects in $\DC(X)$ as D-branes. The D-branes that are
truly physical must be $\Pi$-stable \cite{DFR:stab} (which, from now
on, we refer to simply as ``stable''). Unlike the derived category
data, the stability condition depends on the complexified K\"ahler
form $B+iJ\in H^2(X,\C)$ which varies as one flops between $X$ and
$X'$. As is well-known, and we review in section \ref{s:flop},
imposing stability eliminates just the right set of candidate 0-branes
and the remaining set of D-branes nicely form a moduli space
corresponding to the target space $X$, $X'$, or the singular space at
the ``middle'' of the flop.

The purpose of this paper is to explore whether such a pretty picture
exists for more subtle examples. The answer, we will see, is much less
clear-cut. A general framework for providing numerous examples is
given by the gauged linear sigma-model of \cite{W:phase}. In this
context one has various phases one may probe by varying $B+iJ$. It is
natural to ask if the geometry of other phases can be seen by D-brane
probes. 

The geometry of the other phases was previously explored by considering
the complexified K\"ahler form $B+iJ$ itself. One may view this as
measuring the areas of complex curves in $X$.
In
\cite{AGM:sd,me:min-d} the value of $B+iJ$ was analytically continued
using mirror symmetry from the large radius limit into other
phases. The results for these linear sigma model examples suggest the
following rules:
\begin{enumerate}
\item The smooth \CY\ phases are as expected.
\item Geometric but singular phases such as orbifolds exhibit zero
  sizes in a limit. That is, if we move in the moduli space to blow an
  exceptional divisor down as far as possible while making the rest of
  the \CY\ threefold infinitely large, we measure areas of curves in
  the exceptional set as zero. If $X$ remains finite size, studied
  examples imply that it is impossible to shrink the exceptional
  divisors down to zero size.
\item Non-geometric phases containing Landau-Ginzburg theories or
  fibrations are pushed properly into the K\"ahler cone of a smooth
  phase. That is, no zero sizes are measured.
\end{enumerate}
In the above analysis some choices of branch cuts need to be made to
make these statements. We will discuss this further below.  It was
shown in \cite{W:MF} that an M-theory picture of measuring sizes gives
the same result.

We will see in section \ref{ss:orb} that the 0-brane probe picture of
the orbifold coincides perfectly with the above. Indeed, the stability
of the 0-brane is tied precisely to the value of $B+iJ$ in such a way
that the exceptional set collapses to a point precisely when the
corresponding component of $J$ vanishes.

\begin{figure}
\begin{center}
\includegraphics[width=140mm]{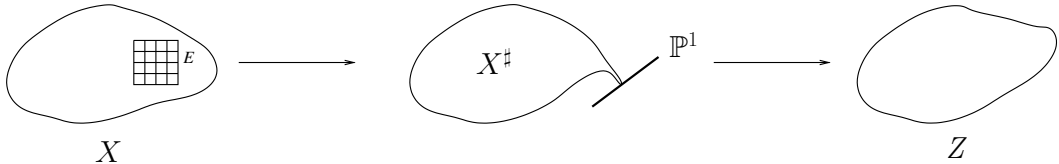}
\end{center}
\caption{An exoflop extremal transition.} \label{fig:et}
\end{figure}

A much more interesting example concerns the ``exoflop''. In this
case, a phase change is achieved by shrinking down some divisor and
then continuing (apparently) beyond the wall of the K\"ahler cone
causing a $\P^1$ to protrude {\em outside\/} the \CY\ threefold. The
phase is considered non-geometric since this $\P^1$ has
Landau--Ginzburg theory fibres over it. It is natural to ask if the
0-brane probe sees this $\P^1$ geometry. Another reason an exoflop
example is of interest is that it is commonly associated to an
extremal transition to a topologically distinct \CY\ threefold
$Z$. That is, when a smooth threefold $X$ is exoflopped by shrinking
down a divisor $E$, it typically becomes a union $X^\sharp\cup\P^1$,
where $X^\sharp$ is a singular threefold touching the external $\P^1$
at its singularity. The space $X^\sharp$ then admits a deformation of
complex structure resulting in a smooth space $Z$. We depict this
process in figure \ref{fig:et}.  This exoflop extremal transition is
the transition that connects the web of hypersurfaces in toric
varieties as studied in \cite{ACJM:srch}.

An extremal transition is associated to massless D-branes as discussed
in \cite{GMS:con}. In the case of an exoflop, this actually happens in
the exoflop {\em limit} rather than in the wall between phases. This
leads to a conundrum. In the limit one might na\"\i vely think that the
$\P^1$ sticking out is very large but then it vanishes completely in
the extremal transition.

We will resolve this somewhat by showing that the 0-brane doesn't
``see'' the $\P^1$ and so the extremal transition makes more
sense. Having said that, we will further show that, if the threefold
has finite volume, then the divisor $E$ cannot be shrunk down to a
point.  Thus the extremal transition still seems very peculiar from
the point of the view of the 0-brane moduli space. The geometry, as
seen by the 0-brane, behaves discontinuously during the extremal
transition as $E$ jumps from nonzero size to not existing. As
we discuss later, this paradox might be explained by considering
string coupling effects.

Having failed to find the $\P^1$ of the exoflop we look at a simpler
hybrid model in section \ref{s:P1}. Here we find a candidate D-brane
by using the technology of matrix factorizations. Its moduli space
appears to be the $\P^1$ base of a \LG\ fibration. It is interesting
that this D-brane is only stable when one is close to the hybrid
phase, i.e., it is unstable in the large radius limit. The charge of
this D-brane is not that of a 0-brane. Therefore to probe the geometry
of hybrid models we need to look further than 0-branes.

It should be emphasized that what one calls ``the'' geometry of a
hybrid phase should be regarded as a pretty subjective statement. One
can have different criteria for measuring the geometry. In this
paper we see discrepancies between the geometry as seen by the
critical points of the superpotential and the moduli space of
0-branes. There are other interesting ways of interpreting the
geometry such as \cite{Caldararu:2007tc} (which has some overlap with
section \ref{s:P1}).


\section{The Flop} \label{s:flop}

In this section we review the well-known case of the flop as discussed in
\cite{Brig:flop,Doug:DC,me:point}. Let $Y$ denote the singular
conifold $xy=zw$ in $\C^4$ which may also be considered the quotient
of $\C^4$ with coordinates $(u_0,u_1,v_0,v_1)$ under the $\C^*$ action
$(u_0,u_1,v_0,v_1)\mapsto(\lambda u_0,\lambda u_1, \lambda^{-1}
v_0,\lambda^{-1} v_1)$.  Let $S=k[u_0,u_1,v_0,v_1]$ be the
singly-graded homogeneous coordinate ring in the sense of Cox
\cite{Cox:} with these weights.  One may identify $x=u_0v_0, y=u_1v_1,
z=u_0v_1, w=u_1v_0$.

Let $X$ be a noncompact small resolution of the conifold given as the
total space of the bundle $\O(-1)\oplus\O(-1)$ over $\P^1$.  The
homogeneous coordinates of $\P^1$ are given by $[u_0,u_1]$ and the
fibre directions are given by $v_0$ and $v_1$.  Let $C\cong\P^1$
denote the zero section, $v_0=v_1=0$ of this bundle.  There are
numerous ways to deduce the result (essentially due to \cite{Bei:res}
and first manifesting itself in the physics literature in
\cite{KW:coni}) that $\DC(X)$ is equivalent to the bounded derived
category of finitely-generated representations of a quiver. Here we
follow possibly the quickest way \cite{me:toricD}. We find a tilting
object
\begin{equation}
   T = \cE_0\oplus\cE_1\oplus\ldots
\end{equation}
as a sum of line bundles in $\DC(X)$. The endomorphism ring $\End(T)$
is then the path algebra (with relations) of a quiver
$Q^{\mathrm{op}}$ and $\DC(X)$ is equivalent to the bounded derived
category of finitely-generated representations of $Q$ (with relations).
We refer to \cite{me:toricD} for details.\footnote{Rather than use the
  opposite quiver, one can equivalently use {\em right\/} $A$-modules.}
In our case, for the flop,
we put
\begin{equation}
  T = \O\oplus\O(1),
\end{equation}
where $\O(n)$ is the sheaf associated to the $S$-module $S(n)$. If
$\pi:X\to\P^1$ represents the fibration, then $\O(n)$ is the
noncompactly supported sheaf $\pi^*\O_{\P^1}(n)$. Associating the node
$v_i$ to the summand $\O(i)$ in the tilting sheaf we obtain the
quiver:
\begin{equation}
\begin{xy} <1.0mm,0mm>:
  (0,0)*{\circ}="a",(30,0)*{\circ}="b",
  (0,-4)*{v_0},(30,-4)*{v_1}
  \ar@{->}@/^4mm/|{v_0} "a";"b"
  \ar@{->}@/^2mm/|{v_1} "a";"b"
  \ar@{->}@/^2mm/|{u_0} "b";"a"
  \ar@{->}@/^4mm/|{u_1} "b";"a"
\end{xy}  \label{eq:conquiv}
\end{equation}
with relations given by the superpotential
$\Tr(u_0v_0u_1v_1-u_0v_1u_1v_0)$.

In general, let $A$ be the opposite algebra of $\End(T)$ and let $e_i$
be the idempotent element of the path algebra given by zero length
paths at node $v_i$. Then $P_i=Ae_i$ is the projective $A$-module
given by all paths (with relations) starting at node $v_i$. The
equivalence between $\DC(X)$ and $\DC(A\textrm{-mod})$ sends a
summand $\cE_i$ of the tilting object to $P_i$. 

Because $X$ is a {\em noncompact\/} \CY\ threefold, the K-theory class
of a 0-brane in $\DC(X)$ as we have described it is actually
zero. Therefore we need to be careful about what we really mean by the
``class'' of a 0-brane. If we consider the derived category of {\em
  compactly supported\/} coherent sheaves on $X$ then this is
equivalent to the derived category of {\em finite-dimensional\/}
quiver representations. See \cite{Bridge:Z3,Berg:compact} for a
further discussion of this.  Let $V$ be a finite dimensional
representation of the the quiver $Q$ (or a complex of representations
with this as the only nonzero entry). In the derived category of
finite-dimensional representations of $Q$, the K-theory charge of $V$
is simply given by its dimension vector $(d_0,d_1,\ldots)$, where
$d_i=\dim e_iV$. We will therefore use this as the characterization of
being in the ``class'' of a 0-brane. 

We will assume that the object in $\DC(A\textrm{-mod})$ corresponding
to a skyscraper sheaf is a complex with only one entry, i.e., it can be
represented by a single quiver. This can be proven explicitly in each
case we consider.  Let $V_p$ denote the quiver representation
associated to a 0-brane, i.e., the skyscraper sheaf $\O_p$ of a point
and suppose it has dimension vector $(d_0,d_1,\ldots)$.  As was shown
in \cite{AM:delP},
\begin{equation}
\begin{split}
  \dim(\Hom(P_i,V_p)) &= d_i\\
     &= \dim(\Hom(\cE_i,\O_p))\\
     &= \rank(\cE_i)\\
     &= 1.
\end{split}
\end{equation}
That is, the skyscraper sheaf corresponds to a quiver representation
with dimension vector $(1,1,\ldots)$.

\subsection{Moduli Space of Quiver Representations} \label{ss:mod0}

In general the analysis of the moduli space of quiver representations
can be fairly technical but has been studied in detail in, for
example, \cite{SI:II,DGM:Dorb}, and using dimer language
\cite{Franco:2006gc,IshUe:dimmod}. By viewing the quiver in terms of
the endomorphism ring of a tilting object $T$ we can find a very easy
method that suffices for the examples in this paper. The idea of using
tilting objects to compute the moduli space was discussed in detail in
\cite{BergP:modq}.

A quiver representation having the dimension vector $(1,1,\ldots)$ is
specified by attaching a single number to each arrow.
There is a class of isomorphisms acting within the set of such
representation by acting with an element of $\C^*$ on each node. This
leads to a torus action on the parameters associated to each
arrow. The arrows on the quiver (\ref{eq:conquiv}) are labeled by
homogeneous coordinates from the identification with $\End(T)$.  (In
general, the arrows will be associated with monomials in the
homogeneous coordinates.) It is easy to see that the $(\C^*)^r$ action
coming from the isomorphisms coincides with the toric
$(\C^*)^r$-action on the homogeneous coordinates.  
Furthermore, such labeling of the arrows in the quiver obviously
satisfies any superpotential relationship since such constraints can be
derived from the structure of $\End(T)$ in the first place.
{\em It is
  therefore natural to identify the homogeneous coordinates of the
  point $p$ with the labels in the arrows of the quiver representation
  $V_p$.}

It is worth emphasizing this point as it is key to the constructions
in the paper. The labels on the quiver (\ref{eq:conquiv}) were first
obtained from maps between the tilting summands given by
multiplication by certain monomials in the homogeneous coordinates. But
now we use these same labels for a quiver representation associated to
a skyscraper sheaf of a point with these homogeneous coordinates. This
parametrization of the skyscraper sheaves allows us to construct a
moduli space naturally.

Of course, to construct a nice moduli space for our skyscraper sheaves
we need to remove some affine subvariety before performing the
$(\C^*)^r$ quotient on the homogeneous coordinates. The D-brane
stability condition amounts to a $\theta$-stability condition on the
quiver in the sense of King \cite{King:th} (as discussed in
\cite{DFR:orbifold}) and so removes a suitable subvariety.

It should be emphasized that the general problem of constructing
anything like a moduli ``space'' of stable objects is a subtle
problem. To be rigorous one is forced into using the language of
abstract stacks \cite{Lieblich:modstck}. Even in the large radius
limit it is difficult to prove that the set of 0-branes forms a nice
moduli space \cite{Bayer:poly}. Here we are able to use the language
of quivers in all our examples to explicitly construct the moduli
spaces in question.

We almost have a particularly cheap proof of the fact that the moduli
space of 0-branes on a toric variety $X$ is $X$ itself. What we should
really infer is that, for suitable stability conditions, $X$ is a
subspace of the moduli space of 0-branes. In general one can have
multiple components of the moduli of 0-branes of which only one need
be $X$. This happens, for example, for the case
$\C^3/(\Z_2\oplus\Z_2)$ as studied in \cite{MP:AdS}. In such cases,
for the spurious extra components, one may parametrize the arrows in a
way inconsistent with the labeling by homogeneous coordinates while
still obeying the superpotential constraints. This does not happen for
any example in this paper. For a more careful treatment we refer to
\cite{BergP:modq}.

\subsection{The Moduli Space from Stability Conditions}

Associating to zero the arrows labeled by $v_0$ and $v_1$ in the quiver
(\ref{eq:conquiv}) gives a representation of the Beilinson quiver for
$\P^1$ \cite{Bei:res}. This corresponds to D-branes supported on
$C$. That is, objects in $i_*\DC(\P^1)$ where $i:C\to X$ is the
inclusion map. Accordingly, 0-branes lying on $C$ will have $v_0=v_1=0$.

Let $(d_0,d_1)$ be the dimension vector of a representation of the
quiver (\ref{eq:conquiv}). The simple one-dimensional representations have a
correspondence:
\begin{equation}
\begin{split}
(1,0)&: \O_C\\
(0,1)&: \O_C(-1)[1].
\end{split}
\end{equation}

The complexified K\"ahler form in this noncompact \CY\ space is
parametrized by the complex number $\int_C(B+iJ)$. By abuse of
notation we refer to this complex number as $B+iJ$.  The central
charge, $Z$, of the D-brane $\O_C$ (and of any shift $\O_C[n]$) has a
simple zero at the origin in the $B+iJ$ plane while $Z(\O_C(-1))$ has a
simple zero at $B+iJ=-1$.

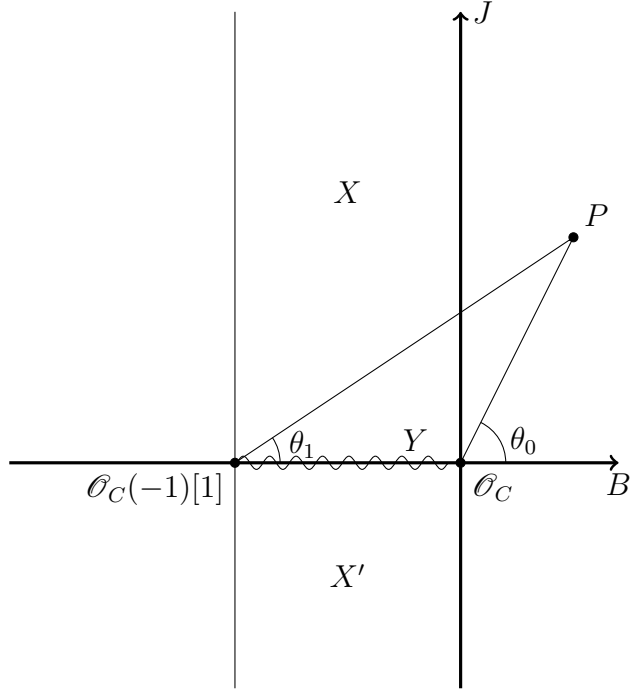
\begin{figure}
\begin{center}
\begin{tikzpicture}[scale=3]
  \draw[very thick,->] (-2,0) -- (0.7,0);
  \draw[very thick,->] (0,-1) -- (0,2) node [anchor=west] {$J$};
  \draw (-1,-1) -- (-1,2);
  \draw (0.7, 0) node [anchor=north] {$B$};
  \draw (0,0) -- (0.5,1) coordinate (P) node [anchor=south west] {$P$};
  \filldraw (P) circle (0.2mm) -- (-1,0);
  \draw (2mm,0) arc (0:63.43:2mm);
  \draw (0.28,0.1) node {$\theta_0$};
  \draw (-8mm,0) arc (0:33.69:2mm);
  \draw (-0.7,0.08) node {$\theta_1$};
  \filldraw (0,0) circle (0.2mm);
  \filldraw (-1,0) circle (0.2mm);
  \draw (0,0) node [anchor=north west] {$\O_C$};
  \draw (-1,0) node [anchor=north east] {$\O_C(-1)[1]$};
  \draw (-0.5,1.2) node {$X$};
  \draw (-0.5,-0.5) node {$X'$};
  \draw[decorate,decoration=snake] (-1,0) -- (0,0);
  \draw (-0.2,0.1) node {$Y$};
\end{tikzpicture}
\end{center}
\caption{The $B+iJ$ plane for the flop.} \label{fig:flop}
\end{figure}

In figure~\ref{fig:flop} we show this $B+iJ$ plane with the above
points labeled by their associated massless D-branes. Now consider
the triangle
\begin{equation}
\xymatrix@C=2mm{\O_C(-1)\ar[rr]^-f&&\O_C\ar[dl]\\
   &\O_p\ar[ul]|{[1]}}
\end{equation}
where $\O_p$ is the 0-brane, i.e., skyscraper sheaf associated to a
point $p\in C$, the location of which is determined by the map $f$.

For $\Pi$-stability we associate a choice of ``phase'' of any object $\cE$:
\begin{equation}
  \xi(\cE) = \frac1\pi\arg Z(\cE),
\end{equation}
(See \cite{me:TASI-D} and references therein for a review.)
With respect to this triangle, $\O_p$ is stable if and only if
\begin{equation}
\xi(\O_C) - \xi(\O_C(-1)) =\frac{\theta_0-\theta_1}{\pi}\leq 1.
\end{equation}
In other words, $\O_p$ decays as it passes below the line segment of
the $B$-axis between the two massless D-brane points in
figure~\ref{fig:flop}.

Furthermore, if one does pass into the region of negative $J$ by
passing across this line segment, we have a triangle
\begin{equation}
\xymatrix@C=2mm{\O_C[-1]\ar[rr]^-g&&\O_C(-1)[1]\ar[dl]\\
   &D_g\ar[ul]|{[1]}}
\end{equation}
defining objects $D_g$ which become stable. These new objects $D_g$
can be interpreted as skyscraper sheaves of points on the {\em
  flopped\/} curve $C'\subset X'$ \cite{Brig:flop}.

Note that {\em proving\/} an object is $\Pi$-stable is awkward as the
logic is somewhat circular. The phases of objects in a distinguished
triangle can only imply a decay of an object associated with one
vertex of the triangle if the decay products themselves are stable.
Now we know that $\O_C$ is definitely stable where it becomes massless
at $B+iJ=0$ since it causes divergences in the 2-dimensional
worldsheet field theory. Similarly $\O_C(-1)$ is most assuredly stable
at the other massless D-brane point in figure~\ref{fig:flop}. Thus, it
does not seem too much of a leap of faith to assume that these two
D-branes are both stable along the line segment joining these two
points making our analysis is correct.

In this case one can rigorously prove these stability assertions by
appealing to Bridgeland's definition of stability \cite{Brg:stab} (see
also \cite{Bergman:2007xb}). Here one reduces the stability statements
to an {\em abelian\/} category given as the heart of a t-structure of
$\DC(X)$. This is most easily done when the phases of objects ``line
up'' in a specific way. In our case, this abelian category $\cA$ is
the category of representations of the quiver (\ref{eq:conquiv}).
Define a new phase
\begin{equation}
  \phi(A) = \xi(A) + \ff12,
\end{equation}
so that $0<\phi(A)<1$ for all objects in $\cA$ so long as we remain
between the lines $B=0$ and $B=-1$ in figure~\ref{fig:flop}.
So long as we obey this condition, all stable D-branes are objects
in $\cA$ (or their translates).
An object $A$ in $\cA$ is {\em semistable\/} so long as
$\phi(E)\leq\phi(A)$ for any subobject $E\subset A$ and stable so long
as $\phi(E)<\phi(A)$. The objects
corresponding to $\O_C$ and $\O_C(-1)[1]$ are then obviously stable as
the dimension vector implies they have no nontrivial subobject.

Consider
an object in $\cA$ with dimension vector $(m,n)$. This has phase
\begin{equation}
  \phi(m,n) = \frac1{\pi}\arg i(nt+n-mt),
\end{equation}
where $t=B+iJ$.  
Now the 0-brane $\O_p$, $p\in C$ is a quiver in the
class $(1,1)$ and is stable according to the following rules:
\begin{itemize}
\item If $(u_0,u_1)\neq(0,0)$ and $(v_0,v_1)\neq(0,0)$, $\O_p$ has no
  subobject and is thus always stable.
\item If $(u_0,u_1)=(0,0)$ then $\O_C(-1)[1]$ is a subobject with dimension
  vector $(0,1)$. This renders this 0-brane unstable if $\Im(t)>0$. So,
  if $J>0$ we may regard $[u_0,u_1]$ as homogeneous coordinates giving
  the location of the 0-brane on $C$.
\item If $(v_0,v_1)=(0,0)$ then $\O_C$ is a subobject with dimension
  vector $(1,0)$. This renders this 0-brane unstable if $\Im(t)<0$. So,
  if $J<0$ we may regard $[v_0,v_1]$ as homogeneous coordinates giving
  the location of the 0-brane on the flopped curve $C'$.
\end{itemize}

So far we have produced the expected moduli space of {\em stable\/}
skyscraper sheaves when $J\neq0$.  Now consider the case where
$\theta_1-\theta_0=\pi$, i.e., we lie on the line segment highlighted
by the wiggly line in figure~\ref{fig:flop}. Some quiver
representations in the class $(1,1)$ are now semistable with respect
to decay into $(0,1)$ and $(1,0)$ but not strictly stable.

We claim one should use the notion of polystability. We will say that
an object in $\DC(X)$ is {\em polystable\/} if it is a direct sum of
strictly stable objects (i.e., strict inequalities are used in
stability conditions). This notion is copied from the study of
Hermitian--Yang--Mills connections on vector bundles. A holomorphic
vector bundle admits an Hermitian--Yang--Mills connection if and only
if it is polystable \cite{Don:YM,UY:YM}.

The reason why polystability is the correct notion for D-branes comes
from the classical notion of mutually BPS solitons. It is well-known
that there is no force between two BPS solitons. Thus, the combined
classical state of two mutually BPS states is simply their direct sum.

A quiver representation with dimension vector $(1,1)$ such that $u_i$
are both zero or $v_i$ are both zero is not strictly stable with
respect to decay into $(0,1)\oplus(1,0)$. The only polystable object in
this collection then consists of the direct sum where all four maps
$(u_0,u_1,v_0,v_1)$ are set to zero. That is, the origin of the
conifold $x=y=z=w=0$ is counted precisely once. 

We have therefore shown that, when $\theta_1-\theta_0=\pi$, the moduli
space of quivers with dimension vector $(1,1)$ is the conifold $Y$. By
using polystability, the moduli space of skyscraper sheaves yields
$X$, $X'$ or $Y$ precisely, according to figure~\ref{fig:flop}.


\section{Shrinking Divisors}  \label{s:orb}

In this section we consider the case of varying $B+iJ$ so as to shrink
a divisor $E$ down to a point. Let's first give a very rough idea of how
this differs from the flop case of the previous section.

Consider the short exact sequence
\begin{equation}
\xymatrix@1{0\ar[r]&\cI_p\ar[r]&\O_E\ar[r]&\O_p\ar[r]&0,} \label{eq:dp1}
\end{equation}
where $\O_p$ is a 0-brane of a point $p\in E$ and $\cI_p$ is the ideal
sheaf of $p$ in $E$ extended by zero. At large radius limit we know
the phase of a sheaf $\cE$ is given by (see, for example,
\cite{me:TASI-D})
\begin{equation}
  \xi(\cE) = -\ff12\dim\supp(\cE).  \label{eq:dimsup}
\end{equation}
So $\xi(\O_E)=-1$ and $\xi(O_p)=0$. Now suppose, as is typical for
\CY\ threefolds from \cite{Horj:EZ,AKH:m0}, we have a point in the
moduli space of conformal field theories where $Z(E)$ acquires a
simple zero. Put this point at the origin of the $B+iJ$ moduli space
as we did for the flop in figure \ref{fig:flop}. Now assume, again as
is typical, that the phase $\xi(\O_E)$ remains constant as we follow
the $J$-axis down from infinity.  In order to destabilize a point
$\O_p$ via a decay associated to (\ref{eq:dp1}) the angle $\theta_0$
in figure \ref{fig:flop} would be expected to be $3\pi/2$ rather that
the $\pi$ seen for the flop (where the dimension of the support was
only 1). It would appear that we get the 0-brane to decay by moving
all the way to the limit of the phase in the lower half-plane of the
figure. This gives a good picture of what actually happens. Rather
than subspaces collapsing to a point on a phase {\em boundary\/} as
happened for the flop, an exceptional divisor collapses at a phase
{\em limit}.

\subsection{An orbifold} \label{ss:orb}

The following is, in some sense, the simplest example of a compact
\CY\ threefold with an orbifold singularity. Let $\cA$ be a collection
of 7 points in $\R^5$ with coordinates:
\begin{equation}
\begin{array}{c|ccccc}
x_0&1&\pz0&\pz0&\pz0&\pz1\\
x_1&1&\pz0&\pz0&\pz1&\pz0\\
x_2&1&\pz0&\pz1&\pz0&\pz0\\
x_3&1&\pz1&\pz0&\pz0&\pz0\\
x_4&1&-1&-1&-1&-3\\
e&1&\pz0&\pz0&\pz0&-1\\
p&1&\pz0&\pz0&\pz0&\pz0
\end{array}
\end{equation}
Let $\Sigma$ be a fan over a triangulation of this point set.  We can
associate a noncompact toric variety $V_\Sigma$ in the usual way to this
data. Let the homogeneous coordinates be named as in the left column
of the table above. Now define a function on $V_\Sigma$:
\begin{equation}
W = p\left(x_0^2(x_1+x_2+x_3+x_4) +
  e^2(x_1^7+x_2^7+x_3^7+x_4^7)\right).
\end{equation}
Let $X_\Sigma$ be the critical point set of this function.  Different
triangulations of $\cA$ lead to different possibilities (or
``phases'') for $X_\Sigma$. It is a simple matter to establish that
there are four phases similar to the analysis in
\cite{CDFKM:I,CFKM:II,AGM:I}
\begin{enumerate}
\item The orbifold phase: $X_\Sigma$ is a hypersurface of degree 7 in the
  weighted projective space $\P^4_{\{3,1,1,1,1\}}$ with homogeneous
  coordinates $[x_0,\ldots,x_4]$. There is an orbifold locally of the
  form $\C^3/\Z_3$ at $p=x_1=\ldots=x_4=0$.
\item The \CY\ phase: The above orbifold point is blown up with an
  exceptional divisor $E\cong\P^2$ given by $p=e=0$.
\item The Landau--Ginzburg phase: $X_\Sigma$ is a fat point.
\item The hybrid phase: $X_\Sigma$ is a fat $\P^2$.
\end{enumerate}

The convex hull of the point set $\cA$ is a reflexive polytope and we
can construct the mirror following Batyrev \cite{Bat:m}. The mirror to
$X_\Sigma$ is defined by a polynomial whose monomials naturally
correspond to the homogeneous coordinates above via the
monomial-divisor mirror map \cite{AGM:mdmm}. Let us denote the
coefficients of these monomials by $\tilde x_0, \tilde x_1,\ldots,
\tilde p$. We can then form ``algebraic coordinates'' on the moduli
space of theories:
\begin{equation}
\begin{split}
   y &= \frac{\tilde x_0\tilde e}{\tilde p^2}\\
   z &= -\frac{\tilde x_1\tilde x_2\tilde x_3\tilde x_4}{\tilde p\tilde e^3}.
\end{split}
\end{equation}
In the smooth \CY\ manifold we may define divisor classes $H$,
corresponding to $x_1=0$, and $L$ corresponding to $x_0=0$. It follows
that $e=0$ corresponds to the class $H-3L$, and we have
intersection numbers
\begin{equation}
  H^3=63,\quad H^2L=21,\quad HL^2 = 7, \quad L^3=2.
\end{equation}

Let us write the complexified K\"ahler form as
\begin{equation}
  B+iJ = hH + lL.
\end{equation}
The monomial-divisor mirror map \cite{AGM:mdmm} (with sign corrections from
\cite{MP:inst}) then yields
\begin{equation}
\begin{split}
   h &= \frac1{2\pi i}\log(y) + O(y,z)\\
   l &= \frac1{2\pi i}\log(z) + O(y,z),
\end{split}
\end{equation}
where $h(y,z)$ and $l(y,z)$ are uniquely determined from these
asymptotics as ratios of solutions of the Picard--Fuchs equations
$\Box_z\Phi(y,z)=\Box_y\Phi(y,z)=0$, where
\begin{equation}
\begin{split}
  \Box_z &= \left(z\frac{\partial}{\partial z}\right)^4 -\\
     &z \left(1+z\frac{\partial}{\partial z}+
               2y\frac{\partial}{\partial y}\right)
     \left(y\frac{\partial}{\partial y}-3z\frac{\partial}{\partial
         z}\right) 
     \left(y\frac{\partial}{\partial y}-3z\frac{\partial}{\partial
         z}-1\right) 
     \left(y\frac{\partial}{\partial y}-3z\frac{\partial}{\partial
         z}-2\right)\\
  \Box_y &= y\frac{\partial}{\partial y}
     \left(y\frac{\partial}{\partial y}-3z\frac{\partial}{\partial z}\right)
    -y\left(2y\frac{\partial}{\partial y}+z\frac{\partial}{\partial z}
         +1\right) 
    \left(2y\frac{\partial}{\partial y}+z\frac{\partial}{\partial z}
         +2\right).
\end{split} \label{eq:PF1}
\end{equation}

Consider first the limit of the moduli space when $y\to 0$, i.e.,
$\Im(h)\to\infty$.  This corresponds to the overall volume of the \CY\
going to infinity. Let $C$ be a line on $E$, the exceptional
$\P^2$. Then $[E]^2=-3[C]$. It follows that
\begin{equation}
  \int_C B+iJ = -\ff13(hH+lL)(H-3L)^2 = l.
\end{equation}
So $l$, and thus $z$, controls the size of the blow-up. The limit
$y\to 0$ decompactifies the \CY\ threefold while keeping $E$ finite.

Indeed, in this limit we have
\begin{equation}
  \Box_z = \left(z\frac{\partial}{\partial z}\right)\Box'_z,
\end{equation}
where
\begin{equation}
  \Box'_z = \left(z\frac{\partial}{\partial z}\right)^3 +
      27z\left(z\frac{\partial}{\partial z}\right)
       \left(z\frac{\partial}{\partial z}+\ff13\right)
       \left(z\frac{\partial}{\partial z}+\ff23\right).
\end{equation}
which is familiar as the Picard--Fuchs equation for the blow-up of
$\C^3/\Z_3$ \cite{AGM:sd}. For more details we refer to section 7.3 of
\cite{me:TASI-D}. 

In the noncompact limit $y\to0$ let us model the geometry of
$\C^3/Z_3$ and its resolution torically. The resulting quiver
description is very well-known \cite{DFR:orbifold}. Let the
homogeneous coordinate ring be $S=k[t,y_0,y_1,y_2]$ with weights
$(-3,1,1,1)$. We using a tilting object $S\oplus S(1)\oplus
S(2)$. This gives a quiver with three arrows between each node:

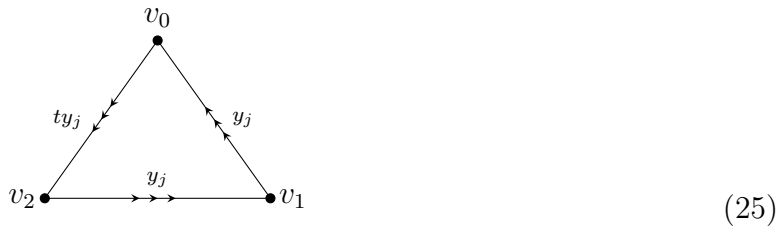
\begin{equation}
\begin{tikzpicture}[scale=1.5,
    decoration={markings,mark=at position 0.42 with {\arrow{stealth}},
                         mark=at position 0.5 with {\arrow{stealth}}, 
                         mark=at position 0.58 with {\arrow{stealth}}}] 
  \filldraw (-1,0) circle (0.4mm);
  \filldraw (1,0) circle (0.4mm);
  \filldraw (0,1.4) circle (0.4mm);
  \draw [postaction={decorate}]  
              (-1,0) -- (1,0) node [midway,above] {$\scriptstyle{y_j}$}; 
  \draw [postaction={decorate}]  
              (1,0) -- (0,1.4) node [midway,right=1mm] 
              {$\scriptstyle{y_j}$}; 
  \draw [postaction={decorate}]  
              (0,1.4) -- (-1,0) node [midway,left=1mm] 
              {$\scriptstyle{ty_j}$}; 
  \draw (0,1.6) node {$v_0$};
  \draw (1.2,0) node {$v_1$};
  \draw (-1.2,0) node {$v_2$};
\end{tikzpicture} \label{eq:qZ3}
\end{equation}
Let $\DC(Q)$ be the derived category of finitely generated
representations of the quiver (\ref{eq:qZ3}) with relations which
follow from the commutativity of the labeling of the arrows. Then
$\DC(Q)$ is a full subcategory of $\DC(X)$. The quotient of $\DC(X)$
by $\DC(Q)$ corresponds to objects that are ``pushed off to infinity''
when we take the decompactification limit $y\to0$. This can be seen by
the fact that $\DC(Q)$ gives a complete description of the local
geometry of the blow-up of $\C^3/\Z_3$.

\begin{figure}
\begin{center}
\begin{tikzpicture}[scale=2,
    decoration={markings,mark=at position 0.64 with {\arrow{stealth}},
                         mark=at position 0.58 with {\arrow{stealth}}, 
                         mark=at position 0.5 with {\arrow{stealth}}}]
  \filldraw [lightgray] (0,0) -- (3.4,2) -- (-3.4,2) -- cycle;
  \filldraw (0,0) circle (0.4mm);
  \filldraw (0,-2) circle (0.4mm);
  \filldraw (1.7,1) circle (0.4mm);
  \filldraw (-1.7,1) circle (0.4mm);
  \draw (1.8,1) node [anchor=west] {$\O_E$ massless};
  \draw [postaction={decorate}] (-1,1.5) -- (-0.6,1.5);  
  \draw [postaction={decorate}] (-0.6,1.5) -- (-0.8,1.8);  
  \draw [postaction={decorate}] (-0.8,1.8) -- (-1,1.5);
  \draw (-1,1.65) node {$\scriptstyle0$};
  \draw (-0.6,1.65) node [anchor=west] {stable};
  \draw [postaction={decorate}] (1,0) -- (1.4,0);  
  \draw [postaction={decorate}] (1.4,0) -- (1.2,0.3);  
  \draw [postaction={decorate}] (1.2,0.3) -- (1,0);
  \draw (1.2,-0.1) node {$\scriptstyle0$};
  \draw (1.4,0.15) node [anchor=west] {stable};
  \draw [thick] (0,0) -- (1.7,1);
  \draw (0,0) node [anchor=east] {$O$};
  \draw (1,0.4) node {$L$};
  \draw (0,-2) node [anchor=north] {$\Omega(1)[1]$ massless};
  \draw (-1.75,1) node [anchor=east] {$\O_E(-1)[2]$ massless};
  \draw (-3,-1) node [draw] {$-i\psi$};
\end{tikzpicture}
\end{center}
\caption{Regions of stability for objects with dimension vector
  $(1,1,1)$.}
    \label{fig:Z3}
\end{figure}
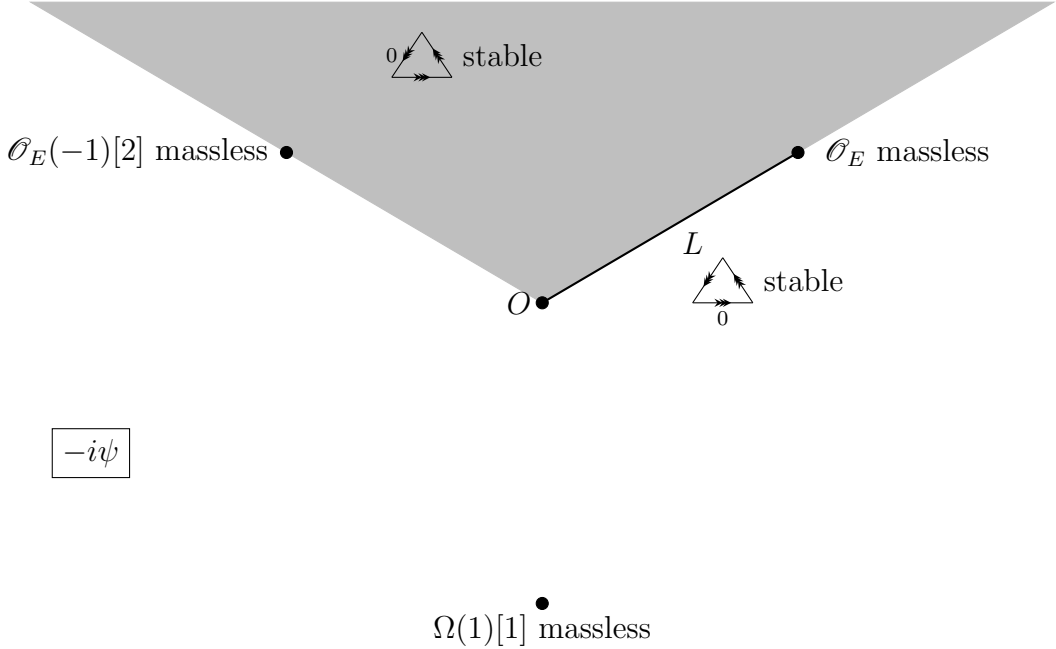

The situation now becomes fairly similar to the conifold in the
previous section. Consider representations of this quiver with
dimension vector $(d_0,d_1,d_2)$. As discussed in
\cite{DFR:orbifold,Asp:theta}, when we take the limit $z\to\infty$,
corresponding to $l\to 0$, the phases, $\xi$, of all representations
become equal. Thus we may describe all stable objects in an open
neighbourhood of this limit by the abelian category of quiver
representations. Setting $z=(3e^{-\pi i}\psi)^{-3}$, we may draw lines
of marginal stability in the $\psi$-plane,

The simple one-dimensional representations correspond to the so-called
``fractional branes'':
\cite{DFR:orbifold,me:TASI-D} 
\begin{equation}
\begin{split}
  (1,0,0)&: \O_E\\
  (0,1,0)&: \Omega_E(1)[1]\\
  (0,0,1)&: \O_E(-1)[2].
\end{split}
\end{equation}
The skyscraper sheaves of a point, $\O_p$, have dimension vectors
$(1,1,1)$. The stability of these objects was studied in detail in
section 7.3.6 of \cite{me:TASI-D}. If the point $p$ lies on $E$ then
$t=0$ which results some of the arrows in (\ref{eq:qZ3}) being
zero. This object is stable to decay against the subobjects $(1,0,0)$
and $(1,1,0)$ only if we lie in the grey region of figure~\ref{fig:Z3}
and the maps do not vanish on the other two sides of the triangle. The
moduli space of these objects is $\P^2$ --- in accordance with these
being points on $E$.

Now consider the (interior of the) line labeled $L$ in
figure~\ref{fig:Z3}. $\O_p$ becomes marginally stable against decay
into the subobject of dimension $(1,0,0)$, i.e., $\O_E$. So, on $L$,
$\O_p$ becomes a polystable object $\O_E\oplus Q$, where $Q$ is an
object of dimension $(0,1,1)$. The maps between node $v_1$ and $v_2$
in $Q$ cannot simultaneously vanish as that would destabilize it
further. Thus, the moduli space of $Q$ is still $\P^2$. Therefore,
{\em even though $\O_p$ degenerates into a direct sum of stable
  objects along $L$, its moduli space remains the same}.  When we pass
beyond $L$, representations with maps between $v_1$ and $v_2$ set to
zero become truly stable and still have moduli space $\P^2$.

At the origin, $O$, when $\psi=0$, if $t=0$ then $\O_p$ is only
polystable as a direct sum of the three fractional branes. So, the
exceptional divisor is replaced by a single point. This agrees with
the picture that $O$ corresponds to the orbifold point.

To summarize, we have proven the following for our noncompact
resolution of $\C^3/\Z_3$, in precise agreement with what we would
expect for our 0-brane probe:
\begin{prop}
  In the neighbourhood of $\psi=0$, where the stable objects always
  lie in the abelian category of representations of the quiver
  (\ref{eq:qZ3}), the moduli space of objects with dimension vector
  $(1,1,1)$ is always the blown-up orbifold, except at $\psi=0$, where
  it is the orbifold itself.
\end{prop}
   
Now let us try to analyze what happens when $y\neq 0$, i.e., we give
our \CY\ a finite volume. We immediately lose apparent contact with
any quiver description of the complete category of D-branes. This is because
the quiver picture arises from a {\em tilting object\/} in the derived
category $\DC(X)$. Serre duality for a compact manifold violates the
tilting condition $\Hom(T,T[n])=0$ for $n\neq 0$ and thus there is no
hope of finding a tilting object.  Indeed there is currently no known
method of rigorously determining the complete stable spectrum of
D-branes for any compact \CY\ threefold.

Having said that, it seems reasonable to suppose that the possible
decay modes of skyscraper sheaves should be unchanged between $y=0$
and small values of $y$. In other words, the objects which were pushed off
to infinity when $y\to0$ are still far away in some sense. So, to
determine whether $\O_p$ is stable we can consider the same decay
channels as above and we need only compute the change in phases,
$\xi$, caused by a nonzero $y$. To phrase this another way, the quiver
representations still provide a picture of a subcategory of $\DC(X)$
and we restrict attention only to this subcategory.

As we saw above, the only issue to study is the question of whether
all skyscraper sheaves $\O_p$, for $x\in E$, correspond to a single
polystable object and thus $E$ appears to collapse to a point. This
happens when and only when the phases of the fractional branes become
equal.

Let $\Phi_1$ be the unique solution of the equations (\ref{eq:PF1})
which behaves as $1+O(y,z)$ near the origin. This is $Z(\O_p)$, the
central charge of the skyscraper sheaf. The central charges of the
fractional branes can be determined as a solution of (\ref{eq:PF1}) by
the method described in \cite{AD:Dstab} which depends on the Chern
characters of the objects in question. Let $\Phi_L$ be the unique
solution that behaves as $\log(z)/2\pi i+O(z)$ and let $\Phi_{L^2}$ be
the unique solution that behaves as $(\log(z)/2\pi i)^2+O(z)$. The
central charge of each fractional brane can be written as
\begin{equation}
  a\Phi_1 + b\Phi_L + c\Phi_{L^2},
\end{equation}
for different real values of $a,b,c$. The only way the phases of all
three fractional branes can align is if
\begin{equation}
\begin{split} 
  \int_C J &= \Im(l)\\
   &= \Im\frac{\Phi_L}{\Phi_1}\\
   &= 0.
\end{split}
\end{equation}

Therefore, we have proven 
\begin{prop}
The 0-brane probe condition for the divisor $E$ to
  collapse to a point requires the classical condition that the
  area of $C$ vanishes.
\end{prop}

This is exactly the problem studied in \cite{me:min-d}. There it was
shown in an example almost the same as this one that, for natural
choices of branch cuts, when $y$ is small and nonzero there is no
choice of $z$ that makes the area of $X$ shrink to zero. The same is
true here as we now show.

Putting $z=(3e^{-\pi i}\psi)^{-3}$ and $y=-3\xi\psi$ we obtain good
coordinates $\xi,\psi$ in the neighbourhood of the orbifold limit
point. One can show the periods have the following form near this
limit point:
\begin{equation}
\begin{split}
  \Phi_1 &= 1 + \ldots\\
  \Phi_L &= \frac{2\pi\sqrt{3}}{\Gamma(\ff23)^3}\psi  +
                \frac{63\Gamma(\ff23)^3}{\pi^2}\xi+ \ldots
\end{split}
\end{equation}
Following the logic of \cite{me:min-d} we impose branch cuts to allow
the above analytic continuation:
\begin{equation}
\begin{split}
  \frac{2\pi}3 < &\arg(\psi) <  \frac{4\pi}3\\
  \frac{2\pi}3 < &\arg(\xi) <  \frac{4\pi}3
\end{split}
\end{equation}
If $\xi$ is nonzero, that is we give a finite size to the \CY\
threefold, we see that we cannot make the area of $C$ vanish and thus
the phases cannot align.

\subsubsection{Discussion of branch cuts}  \label{sss:cuts}

It is certainly true that we may analytically continue the periods in
such a way as to make components of $\Im(J)$ vanish and thus align the
periods to make it look as if some of the 0-branes decay. To do this
we must venture past the branch cuts we imposed above. So what does this
actually mean?

One can view the branch cuts as a boundary of a ``fundamental'' region
for the moduli space. Every conformal field theory is accounted for by
a point within this fundamental region. Going beyond the branch cuts
we encounter points that are equivalent (as conformal field theories)
to points in the fundamental region. 

There is a close link between the idea of analytically continuing
periods and monodromy as explained in \cite{AD:Dstab}. Choose some
basepoint in the moduli space of theories close to the large
radius limit. Consider labeling B-type D-branes in a conformal field
theory by associating them with specific objects in $\DC(X)$. Now
follow a path in the space of $B+iJ$ that extends beyond the branch
cuts to some point $P$. There will be a point $Q$ in the fundamental
region which is equivalent to $P$ if we apply an autoequivalence of
$\DC(X)$ to relabel the D-branes. A path from $P$ to $Q$ is a loop in
the moduli space of theories. Monodromy around these loops results in
the corresponding autoequivalence of $\DC(X)$.

Thus, if we destabilize the 0-brane by going beyond the branch cuts,
we can this as equivalent to destabilizing another D-brane while
staying within the fundamental region. Thus, it is more natural to regard
this as using another D-brane probe rather than the 0-brane. 

The view we propose is that we fix a fundamental region by
choosing cuts to make the 0-brane {\em as stable as possible}.  This
is the obvious choice if we want to retain the closest link to
recognizable geometry. For the noncompact orbifold example above this
still resulted in the 0-branes on $E$ becoming unstable in the limit
that $E$ shrink to a point. For the compact orbifold we were able to
make the points of $E$ always stable. By analytically continuing
outside the fundamental region we discover that other D-branes may
decay but they are not 0-branes --- they are merely related to
0-branes by monodromy.

\subsection{Shrinking a Quadric Surface}  \label{ss:exo}

\subsubsection{An exoflop extremal transition}

Let $\cA$ be a collection
of 8 points in $\R^5$ with coordinates:
\begin{equation}
\begin{array}{c|ccccc}
p&1&\pz0&\pz0&\pz0&\pz0\\
a&1&\pz1&\pz0&\pz0&\pz0\\
b&1&\pz0&\pz1&\pz0&\pz0\\
c&1&-2&-1&\pz0&\pz0\\
d&1&\pz0&\pz0&\pz1&\pz0\\
e&1&\pz0&\pz0&\pz0&\pz1\\
f&1&-1&\pz2&-1&-1\\
g&1&-1&\pz0&\pz0&\pz0
\end{array} \label{eq:exo}
\end{equation}
Let $\Sigma$ be a fan over a triangulation of this point set.  Again
associate a noncompact toric variety $V_\Sigma$ to this data. Let the
homogeneous coordinates be named as in the left column of the table
above. Now consider a lowest-order well-defined generic function on $V_\Sigma$ which
will be of the form (omitting coefficients)
\begin{equation}
W = p\biggl(a^2+g^2\Bigl(b^4\left(d^{10}+e^{10}+f^{10}\right)
     + c^4\left(d^2+e^2+f^2\right)\Bigr)\biggr)+\ldots  \label{eq:Edef}
\end{equation}
Let $X_\Sigma$ be the critical point set of this function. Again, different
triangulations of $\cA$ lead to different phases for $X_\Sigma$. There
are 10 distinct triangulations of $\cA$ but not all of these
correspond to different phases of $X_\Sigma$. We are only concerned
with two phases which we label by the corresponding Stanley--Reisner
ideal $I$:
\begin{itemize}
\item The \CY\ phase: $I=(bc,adef,g)$. $X_\Sigma$ is a smooth
  threefold with $h^{1,1}=2$ and $h^{2,1}=144$ (courtesy of PALP
  \cite{Kreuzer:2002uu}). Let $X$ denote this manifold.
\item The exoflop phase: $I=(pb,bc,acdef,g)$. The critical point set
  $X_\Sigma$ is a reducible variety. One component is a threefold with
  an isolated singularity which we denote $X^\sharp$. The other
  component is a (fat) $\P^1$ which intersects $X^\sharp$ transversely
  at its singular point. Such exoflops were first discussed in \cite{AGM:I}.
\end{itemize}

The convex hull of the points (\ref{eq:exo}) is a reflexive
polytope. If we delete the point corresponding to $b$, we change the
convex hull but it is still reflexive. This latter polytope is
associated to a \CY\ threefold $Z$ which is isomorphic to a resolved
hypersurface of degree 10 in $\P^4_{\{5,2,1,1,1\}}$. It has Hodge
numbers $h^{1,1}=1$ and $h^{2,1}=145$. If we deform the complex
structure of $Z$ so that the only monomials which appear in the
defining equation correspond to those of (\ref{eq:Edef}) then $Z$
becomes singular and isomorphic to $X^\sharp$.

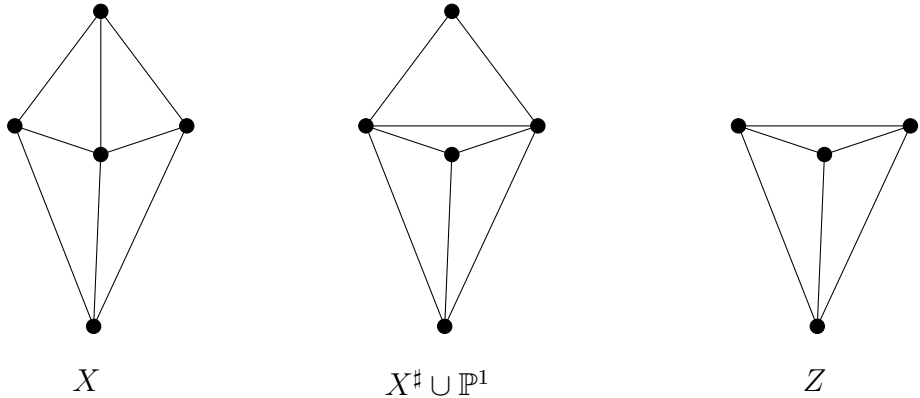
\begin{figure}
\begin{center}
\begin{tikzpicture}[y=-1cm,scale=0.6]

\path[draw=black,fill=white!0!black] (3.96875,8.89) circle (0.15875cm);
\path[draw=black,fill=white!0!black] (6.0325,4.445) circle (0.15875cm);
\path[draw=black,fill=white!0!black] (2.2225,4.445) circle (0.15875cm);
\path[draw=black,fill=white!0!black] (4.1275,1.905) circle (0.15875cm);
\path[draw=black,fill=white!0!black] (4.1275,5.08) circle (0.15875cm);
\path[draw=black,fill=white!0!black] (11.7475,8.89) circle (0.15875cm);
\path[draw=black,fill=white!0!black] (13.81125,4.445) circle (0.15875cm);
\path[draw=black,fill=white!0!black] (10.00125,4.445) circle (0.15875cm);
\path[draw=black,fill=white!0!black] (11.90625,1.905) circle (0.15875cm);
\path[draw=black,fill=white!0!black] (11.90625,5.08) circle (0.15875cm);
\path[draw=black,fill=white!0!black] (20.0025,8.89) circle (0.15875cm);
\path[draw=black,fill=white!0!black] (22.06625,4.445) circle (0.15875cm);
\path[draw=black,fill=white!0!black] (18.25625,4.445) circle (0.15875cm);
\path[draw=black,fill=white!0!black] (20.16125,5.08) circle (0.15875cm);
\draw[black] (4.1275,1.905) -- (2.2225,4.445) -- (3.96875,8.89) -- (6.0325,4.445) -- cycle;
\draw[black] (11.90625,1.905) -- (10.00125,4.445) -- (11.7475,8.89) -- (13.81125,4.445) -- cycle;
\draw[black] (4.1275,1.905) -- (4.1275,5.08) -- (3.96875,8.89);
\draw[black] (2.2225,4.445) -- (4.1275,5.08) -- (6.0325,4.445);
\draw[black] (10.00125,4.445) -- (13.81125,4.445) -- (11.90625,5.08) -- (11.7475,8.89);
\draw[black] (10.00125,4.445) -- (11.90625,5.08);
\draw[black] (18.25625,4.445) -- (20.0025,8.89) -- (22.06625,4.445) -- (18.25625,4.445) -- (20.16125,5.08) -- (20.0025,8.89);
\draw[black] (20.16125,5.08) -- (22.06625,4.445);
\path (3.81,9.6) node [anchor=north] {$X$};
\path (11.58875,9.6) node[anchor=north] {$X^\sharp\cup\P^1$};
\path (19.94375,9.6) node[anchor=north] {$Z$};

\end{tikzpicture}%
\end{center}
\caption{An extremal transition via an exoflop} \label{fig:exo}
\end{figure}

The passage from $X$ to $Z$ via $X^\sharp$ is an extremal transition
exactly of the type discussed for connecting the web of Batyrev-like
\CY\ threefolds in \cite{ACJM:srch}. The change in triangulation is
shown schematically in figure~\ref{fig:exo}.  Note that this geometry
shows that exoflops (rather than conifolds associated to regular
flops) are the general mechanism for extremal transitions between
hypersurfaces in toric varieties.

Suppose we are in the \CY\ phase. The subvariety of $V_\Sigma$
obtained by setting $p=b=0$ is a $\P^3$ with homogeneous coordinates
$[a,d,e,f]$. This intersects the \CY\ $X$ along the smooth quadric surface
\begin{equation}
  a^2 + d^2 + e^2 + f^2 = 0,
\end{equation}
which we denote $E$. In passing between the \CY\ phase and the exoflop
phase we shrink the $\P^3$ down in $V_\Sigma$ and thus we shrink $E$
down. Then, as the exoflop continues, a $\P^1$ grows out transverse to
the \CY\ threefold. At least, this is the geometric description
according to the critical point set $X_\Sigma$. What we would like to
know is whether the moduli space of skyscraper sheaves agrees with
this. We will see that it does not.

A smooth quadric surface is always isomorphic to $\P^1\times
\P^1$. That said, when $\P^1\times \P^1$ is embedded in $\P^3$ as a
quadric surface, the homology classes of the two distinct rulings
become equivalent. In other words, the two $\P^1$ families in the
quadric have the same area. The fact that $h^{1,1}(X)=2$ means that
homology classes in $X$ are all inherited from ones in the ambient
toric variety. Thus, the two rulings of the quadric surface are
homologous in $X$.

To describe the intrinsic geometry of $E$
we will initially suppose the two rulings have different homology
classes, but they will have to be equated once $E$ is embedded in the
\CY\ threefold $X$.

\subsubsection{The infinite volume case}

As in the previous section, first let us make the approximation that
all of the \CY\ is large except for the quadric surface $E$. Thus, $X$
looks like the canonical line bundle of the del Pezzo surface $E$. It
is well-known \cite{MP:AdS,CFIKV:,Herz:exc,AM:delP} how to analyze
D-branes in this geometry using quivers.

Let $E\cong\P^1\times\P^1$ have homogeneous coordinates
$[u_0,u_1],[v_0,v_1]$ and let $p$ denote the coordinate in the normal
direction for the embedding of $E$ in the \CY\ threefold. Let
$\O_E(m,n)$ denote a line bundle on $E$ with first Chern class $me_1+ne_2$
where $e_1$ is the 2-form dual to one of the rulings by $\P^1$ and $e_2$
is dual to the other. In our noncompact limit we have a fibration
$\pi:X\to E$. Let $\O(m,n)$ denote $\pi^*\O_E(m,n)$. To describe $\DC(X)$
we need a tilting collection which amounts to much the same thing as
asking for an exceptional collection on $E$. One can find tilting
collections of line bundles for a toric example such as this one by
using the method described in \cite{me:toricD}. There are four choices
of interest:
\begin{enumerate}[a)]
\item  $\O \oplus \O(0,1) \oplus \O(1,0) \oplus \O(1,1)$
\item  $\O \oplus \O(0,1) \oplus \O(-1,0) \oplus \O(1,1)$
\item  $\O \oplus \O(0,-1) \oplus \O(-1,0) \oplus \O(1,1)$
\item  $\O \oplus \O(0,-1) \oplus \O(1,0) \oplus \O(1,1)$
\end{enumerate}
These are all related by a tilting equivalence which is a form of
Seiberg duality \cite{BD:tilt}.

\begin{figure}
\begin{center}
\begin{tikzpicture}[scale=2,
    decoration={markings,mark=at position 0.45 with {\arrow{stealth}},
                         mark=at position 0.55 with
                         {\arrow{stealth}}}]
  \node at (-0.5,2.5) {a)};
  \node at (0,2) (a) {};
  \filldraw (node cs:name=a)+(0,0) circle (0.4mm);
  \filldraw (node cs:name=a)+(1,0) circle (0.4mm);
  \filldraw (node cs:name=a)+(0,1) circle (0.4mm);
  \filldraw (node cs:name=a)+(1,1) circle (0.4mm);
  \draw (node cs:name=a)+(1,0) node [anchor=north west] {$\O_E(-1,0)[1]$}
         [postaction={decorate}] -- +(0,0) node [anchor=north east]
         {$\O_E$} node [midway,anchor=north] {$u_i$};
  \draw (node cs:name=a)+(0,1) node [anchor=south east] {$\O_E(0,-1)[1]$}
         [postaction={decorate}] -- +(0,0)
         node [midway,anchor=east] {$v_i$};
  \draw (node cs:name=a)+(1,1) node [anchor=south west] {$\O_E(-1,-1)[2]$}
         [postaction={decorate}] -- +(1,0)
         node [midway,anchor=west] {$v_i$};
  \draw (node cs:name=a)+(1,1) [postaction={decorate}] -- +(0,1)
         node [midway,anchor=south] {$u_i$};
  \draw (node cs:name=a)+(0,0) [postaction={decorate,decoration=
      {markings,mark=at position 0.50 with {\arrow{stealth}},
                mark=at position 0.60 with {\arrow{stealth}},}}] -- +(1,1)
         node [midway,below right=-2.7pt] {$pu_iv_j$};
  \node at (-0.5,0.5) {c)};
  \node at (0,0) (a) {};
  \filldraw (node cs:name=a)+(0,0) circle (0.4mm);
  \filldraw (node cs:name=a)+(1,0) circle (0.4mm);
  \filldraw (node cs:name=a)+(0,1) circle (0.4mm);
  \filldraw (node cs:name=a)+(1,1) circle (0.4mm);
  \draw (node cs:name=a)+(0,0) node [anchor=north east] {$\cE[1]$} 
             [postaction={decorate}] -- +(1,0)
             node [midway,below] {$u_i$};
  \draw (node cs:name=a)+(0,0) [postaction={decorate}] -- +(0,1)
             node [midway,left] {$v_i$};
  \draw (node cs:name=a)+(1,0) node [anchor=north west] {$\O_E(-1,0)$}
             [postaction={decorate}] -- +(1,1)
             node [midway,right] {$pv_i$};
  \draw (node cs:name=a)+(0,1) node [anchor=south east] {$\O_E(0,-1)$}
             [postaction={decorate}] -- +(1,1)
             node [midway,above] {$pu_i$};
  \draw (node cs:name=a)+(1,1) node [anchor=south west] {$\O_E(-1,-1)[2]$} 
             [postaction={decorate,decoration=
      {markings,mark=at position 0.50 with {\arrow{stealth}},
                mark=at position 0.60 with {\arrow{stealth}},}}] -- +(0,0)
           node [midway,below right=-2pt] {$u_iv_j$};
\node at (3.5,0.5) {d)};
  \node at (4,0) (a) {};
  \filldraw (node cs:name=a)+(0,0) circle (0.4mm);
  \filldraw (node cs:name=a)+(1,0) circle (0.4mm);
  \filldraw (node cs:name=a)+(0,1) circle (0.4mm);
  \filldraw (node cs:name=a)+(1,1) circle (0.4mm);
  \draw (node cs:name=a)+(1,0) node [anchor=north west] {$\O_E(-1,0)[1]$} 
          [postaction={decorate}] -- +(0,0)
             node [midway,below] {$u_i$};
  \draw (node cs:name=a)+(0,0) node [anchor=north east] {$\O_E(0,-2)[1]$}  
          [postaction={decorate}] -- +(0,1)
             node [midway,left] {$v_i$};
  \draw (node cs:name=a)+(1,1) node [anchor=south west] {$\O_E(-1,-1)[2]$}  
          [postaction={decorate}] -- +(1,0)
             node [midway,right] {$v_i$};
  \draw (node cs:name=a)+(0,1) node [anchor=south east] {$\O_E(0,-1)$}   
          [postaction={decorate}] -- +(1,1)
             node [midway,above] {$pu_i$};
  \node at (3.5,2.5) {b)};
  \node at (4,2) (a) {};
  \filldraw (node cs:name=a)+(0,0) circle (0.4mm);
  \filldraw (node cs:name=a)+(1,0) circle (0.4mm);
  \filldraw (node cs:name=a)+(0,1) circle (0.4mm);
  \filldraw (node cs:name=a)+(1,1) circle (0.4mm);
  \draw (node cs:name=a)+(0,0) node [anchor=north east] {$\O_E(-2,0)[1]$} 
          [postaction={decorate}] -- +(1,0)
             node [midway,below] {$u_i$};
  \draw (node cs:name=a)+(0,1) node [anchor=south east] {$\O_E(0,-1)[1]$}  
          [postaction={decorate}] -- +(0,0)
             node [midway,left] {$v_i$};
  \draw (node cs:name=a)+(1,0) node [anchor=north west] {$\O_E(-1,0)$} 
          [postaction={decorate}] -- +(1,1)
             node [midway,right] {$pv_i$};
  \draw (node cs:name=a)+(1,1) node [anchor=south west] {$\O_E(-1,-1)[2]$} 
          [postaction={decorate}] -- +(0,1)
             node [midway,above] {$u_i$};
\end{tikzpicture}
\end{center}
\caption{Quivers associated to a quadric surface.}
    \label{fig:exo1}
\end{figure}
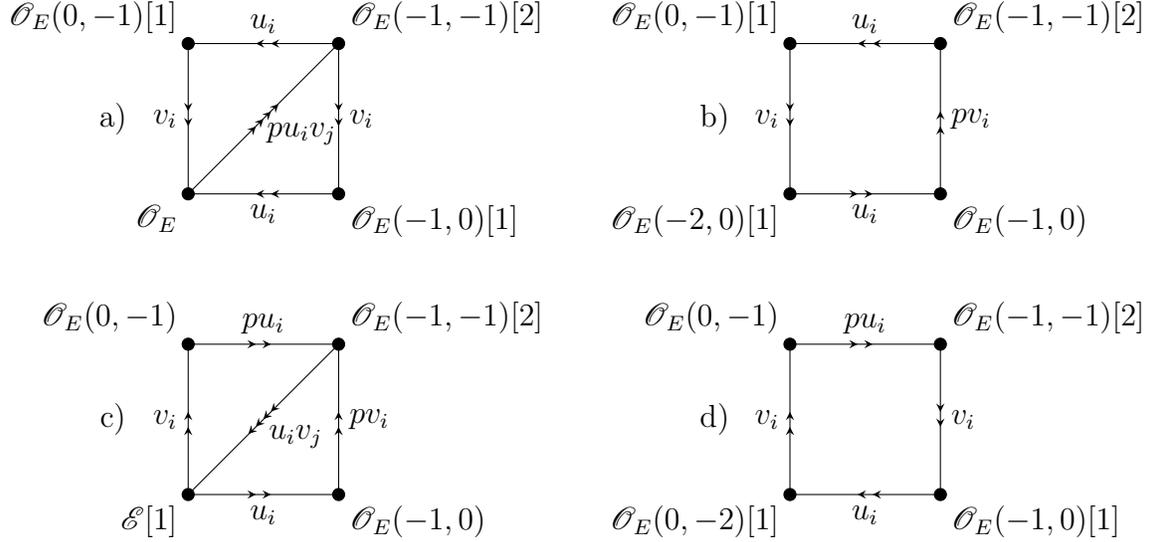

The corresponding quivers are shown in figure
\ref{fig:exo1}. Associated to each node in the quiver is an object (a
``fractional brane'') corresponding to a dimension one representation
of the quiver. We label each vertex in figure~\ref{fig:exo1} by the
corresponding fractional brane. A straightforward method for determining
these fractional branes comes from \cite{me:toricD}. The sheaf $\cE$
in figure \ref{fig:exo1}c) corresponds to a rank 3 vector bundle on
$E$ and is given by the short exact sequence
\begin{equation}
\xymatrix@1@C=14mm{
0\ar[r]&\O_E(-2,-2) \ar[r]^-{\left(\begin{smallmatrix}u_0v_0\\
     u_0v_1\\u_1v_0\\u_1v_1\end{smallmatrix}\right)}&
     \O_E(-1,-1)^{\oplus4}\ar[r]&\cE\ar[r]&0.}
\end{equation}

The maps along the edges of the quiver label the corresponding
homomorphisms between the summands of the tilting collection (with the
arrows reversed) and, as above, these labels will also be associated
to the homogeneous coordinates of the location of a 0-brane. The quiver
relations coming from the superpotential can easily be read off from
this labeling simply from the fact that the coordinates $u_i,v_i,p$
commute.

The analysis of the Picard-Fuchs equation for this system was
performed in \cite{AM:delP}. Let $B+iJ=t_1e_1+t_2e_2$. On the mirror,
we have B-model algebraic coordinates $z_1,z_2$ such that
\begin{equation}
\begin{split}
  t_1 &= \frac1{2\pi i}\log(z_1) + O(z_1,z_2)\\
  t_2 &= \frac1{2\pi i}\log(z_2) + O(z_1,z_2).
\end{split}
\end{equation}
Let us also use the notation $t_{12}$ for the period which
asymptotically goes as
\begin{equation}
  t_{12} = t_1t_2 + O(z_1,z_2).
\end{equation}
There are four linearly independent solutions of the Picard--Fuchs
equation corresponding to the even cohomology of $E$. By using the
techniques of \cite{AGM:sd,me:min-d} we find the following solutions:
\begin{equation}
\begin{split}
\Phi_0(z_1,z_2) &= 1\\
\Phi_1(z_1,z_2) &= t_1 - \ff12 = \frac{1}{2\pi i} \log(e^{-i \pi} z_1) + 
\frac{1}{i \pi} \sum_{(m,n)\neq(0,0)} A_{mn} z_1^m z_2^n\\
\Phi_2(z_1,z_2) &= t_2 - \ff12 = \frac{1}{2\pi i} \log(e^{-i \pi} z_2) + 
\frac{1}{i \pi} \sum_{(m,n)\neq(0,0)} A_{mn} z_1^m z_2^n\\
\Phi_3(z_1,z_2) &= t_{12}-\ff12(t_1+t_2)+\ff14 = 
              -\frac{1}{4\pi^2} \Biggl(\log(e^{-i\pi} z_1) 
\log(e^{-i\pi} z_2)\\
 &\qquad+ \sum_{(m,n)\neq(0,0)} A_{mn} z_1^m z_2^n 
                        \left( \log(e^{-i\pi} z_1) + 
\log(e^{-i\pi} z_2) + \chi_{mn} \right)\Biggr)
\end{split}
\end{equation}
where
\begin{equation}
\begin{split}
A_{mn}    &= \frac{\Gamma(2m + 2n)}{\Gamma(m+1)^2 
\Gamma(n+1)^2}, \nonumber \\
\chi_{mn} &= 4 \Psi(2m+2n) - 2 \Psi(m+1)  -2 \Psi(n+1),
\end{split}
\end{equation}
and $\Psi(z)$ is the usual di-Gamma function, and the sum
$\sum_{(m,n)\neq(0,0)}$ runs over all non-negative $m,n$ with the
exception of $m=n=0$. The power series converge for $
\left|z_1\right| < \ff14$ and $\left|z_2\right| < \ff14$ which
corresponds to the large radius limit phase. We may analytically
continue beyond this phase to obtain the following, where
$y_1=z_2^{-1}z_1$ and $y_2=z_2^{-1}$:
\def\wchi{\widetilde{\chi}}
\begin{equation}
\begin{split}
\Phi_0(y_1,y_2)  &= 1\\
\Phi_1(y_1,y_2)  &= \frac{1}{2\pi i} \log(y_1) + \Phi_2(y_1,y_2)\\ 
\Phi_2(y_1,y_2)  &= -\frac{1}{2\pi i} 
\left(e^{i\pi}y_2\right)^{\frac{1}{2}}
                       \sum_{m,n\geq0} B_{mn} y_1^m y_2^n\\
\Phi_3(y_1,y_2)  &= \frac{1}{12} - \frac{1}{4\pi^2} 
\left(e^{i \pi} y_2\right)^{\frac{1}{2}}
                       \sum_{m,n\geq0} B_{mn} y_1^m y_2^n 
\left(-\log(y_1) + \wchi_{mn} \right),
\end{split} \label{eq:smallr}
\end{equation}
where expressions for $B_{mn}$ and $\wchi_{mn}$ can be found in \cite{AM:delP}.

When this del Pezzo surface is embedded into the \CY\ threefold we
must set $z_1=z_2$. This puts us right on the limit of convergence for 
the power series in (\ref{eq:smallr}). The limit of the exoflop phase
corresponds to $y_1=1$ and $y_2=0$. That is, $\Phi_0=1$,
$\Phi_1=\Phi_2=0$ and $\Phi_3=1/12$. This, in turn implies
$t_1=t_2=1/2$ and $t_{12}=1/3$.

The fact that $t_1=t_2=1/2$ in this limit implies that the K\"ahler
form $J$ is zero and $B$ has component $1/2$ on each ruling. So, using
the criteria of \cite{AGM:sd}, by measuring areas with the K\"ahler form,
$E$ has shrunk down to zero size. Is the same true using the moduli
space of 0-branes?

The central charges of interest are given by
\begin{equation}
  Z(\O_E(-m,-n)) = (\ff76-m-n+mn) + (n-1)t_1 + (m-1)t_2 + t_{12}, \label{eq:Zmn}
\end{equation}
which gives
\begin{equation}
  Z(\O_E(-m,-n)) = \ff12 + mn - \ff12(m+n)
\end{equation}
in the exoflop limit. In particular, we have
\begin{equation}
  Z(\O_E(-1,0)) = Z(\O_E(0,-1)) = 0,
\end{equation}
giving two massless fractional D-branes in each of the quivers in figure
\ref{fig:exo1}. In each case the other two fractional branes have
central charge equal to $1/2$.

When we embed $E$ into $X$ we identify the classes $e_1$ and $e_2$, and
thus $\O_E(-1,0)$ and $\O_E(0,-1)$ have the same Chern class. That
said, they are two distinct objects in the derived category and thus
two distinct D-branes. To see this, note that
$\Hom(\O_E(-1,0),\O_E(0,-1))=0$. 

Stable massless D-branes correspond to singular conformal field
theories and, indeed, one can compute the discriminant
\begin{equation}
  \Delta = 16 y_1^2 - 8 y_1 y_2 + y_2^2 - 32 y_1 -8 y_2 + 16,
\end{equation}
to confirm that $y_1=1,y_2=0$ is singular.

The fact that {\em two\/} D-branes become massless means that we have
an extremal transition, following \cite{GMS:con}. Indeed, it is
precisely two massless states that are required to change the Hodge
numbers from $(2,144)$ to $(1,145)$. 

One should note that the point in moduli space where $z_1=z_2=1/4$ is
also of interest. This is where the discriminant intersects the wall
between the \CY\ phase and the exoflop phase. In this case just a
single D-brane goes massless according to the EZ analysis of
\cite{Horj:EZ,AKH:m0}. Hence there is no extremal transition here;
just a singular conformal field theory. It is worth emphasizing that
this exoflop example is a little unusual compared to simpler examples
in that the limit point of the phase, and not just the phase
transition wall, hits the discriminant.

The quiver associated to the exoflop limit is the one where all the
phases of the simple objects align. Typically this picks out a unique
quiver. This is because tilting transformations of the type studied in
\cite{AM:delP} replace one of the fractional branes by an antibrane
which therefore destroys any alignment of phases. The exception to
this, of course, is when certain fractional branes are massless. Then
the tilted quiver replacing a brane by its anti-brane is just as
valid. In our case all four quivers listed in figure \ref{fig:exo1}
are relevant.

The quiver representation associated to the 0-brane has dimension
vector $(1,1,1,1)$ in all four cases and the labels on the edges
correspond to the required maps in the representation to move the
0-brane around $X$. Thus, to place the 0-brane on $E$ we set any edge
containing a $p$ to zero. The quiver c) in figure \ref{fig:exo1} is of
particular interest in this case. For 0-branes on $E$ we have a short
exact sequence
\def\ss{\scriptstyle}
\begin{equation}
\xymatrix@1{
0\ar[r]&
\begin{tikzpicture}[scale=1,
    decoration={markings,mark=at position 0.45 with {\arrow{stealth}},
                         mark=at position 0.55 with
                         {\arrow{stealth}}}]
  \filldraw (node cs:name=a)+(0,0) circle (0.4mm);
  \filldraw (node cs:name=a)+(1,0) circle (0.4mm);
  \filldraw (node cs:name=a)+(0,1) circle (0.4mm);
  \filldraw (node cs:name=a)+(1,1) circle (0.4mm);
  \draw (node cs:name=a)+(0,0) node [below left] {$\ss1$}
         [postaction={decorate}] -- +(1,0);
  \draw (node cs:name=a)+(0,0) 
         [postaction={decorate}] -- +(0,1) node [above left] {$\ss0$};
  \draw (node cs:name=a)+(1,0) node [below right] {$\ss1$}
         [postaction={decorate}] -- +(1,1) node [above right] {$\ss1$}
         node [midway,right] {$\ss0$};
  \draw (node cs:name=a)+(0,1) [postaction={decorate}] -- +(1,1);
  \draw (node cs:name=a)+(1,1) [postaction={decorate,decoration=
      {markings,mark=at position 0.50 with {\arrow{stealth}},
                mark=at position 0.60 with {\arrow{stealth}},}}] --
            +(0,0);
\end{tikzpicture}
\ar[r]&
\begin{tikzpicture}[scale=1,
    decoration={markings,mark=at position 0.45 with {\arrow{stealth}},
                         mark=at position 0.55 with
                         {\arrow{stealth}}}]
  \filldraw (node cs:name=a)+(0,0) circle (0.4mm);
  \filldraw (node cs:name=a)+(1,0) circle (0.4mm);
  \filldraw (node cs:name=a)+(0,1) circle (0.4mm);
  \filldraw (node cs:name=a)+(1,1) circle (0.4mm);
  \draw (node cs:name=a)+(0,0) node [below left] {$\ss1$}
         [postaction={decorate}] -- +(1,0);
  \draw (node cs:name=a)+(0,0) 
         [postaction={decorate}] -- +(0,1) node [above left] {$\ss1$};
  \draw (node cs:name=a)+(1,0) node [below right] {$\ss1$}
         [postaction={decorate}] -- +(1,1) node [above right] {$\ss1$}
         node [midway,right] {$\ss0$};
  \draw (node cs:name=a)+(0,1) [postaction={decorate}] -- +(1,1)
         node [midway,above] {$\ss0$};
  \draw (node cs:name=a)+(1,1) [postaction={decorate,decoration=
      {markings,mark=at position 0.50 with {\arrow{stealth}},
                mark=at position 0.60 with {\arrow{stealth}},}}] --
            +(0,0);
\end{tikzpicture}
\ar[r]&
\begin{tikzpicture}[scale=1,
    decoration={markings,mark=at position 0.45 with {\arrow{stealth}},
                         mark=at position 0.55 with
                         {\arrow{stealth}}}]
  \filldraw (node cs:name=a)+(0,0) circle (0.4mm);
  \filldraw (node cs:name=a)+(1,0) circle (0.4mm);
  \filldraw (node cs:name=a)+(0,1) circle (0.4mm);
  \filldraw (node cs:name=a)+(1,1) circle (0.4mm);
  \draw (node cs:name=a)+(0,0) node [below left] {$\ss0$}
         [postaction={decorate}] -- +(1,0);
  \draw (node cs:name=a)+(0,0) 
         [postaction={decorate}] -- +(0,1) node [above left] {$\ss1$};
  \draw (node cs:name=a)+(1,0) node [below right] {$\ss0$}
         [postaction={decorate}] -- +(1,1) node [above right] {$\ss0$};
  \draw (node cs:name=a)+(0,1) [postaction={decorate}] -- +(1,1);
  \draw (node cs:name=a)+(1,1) [postaction={decorate,decoration=
      {markings,mark=at position 0.50 with {\arrow{stealth}},
                mark=at position 0.60 with {\arrow{stealth}},}}] --
            +(0,0);
\end{tikzpicture}
\ar[r]&0.}
\end{equation}
This shows how the fractional brane $\O_E(0,-1)$ can break off the
0-brane. Similar $\O_E(-1,0)$ can also be a decay product. 
Now both of these fractional branes are massless at the exoflop point
and have no well-defined phase $\arg(Z)$. The obvious treatment of
this case, using the usual BPS mass rules, is to assert that the
0-brane is marginally stable against decays involving these decay
products. A rigorous argument along these lines is a little delicate
because we are assuming we are in a classical zero-string coupling
limit. In fact, the actual D-brane mass is proportional to $Z/g$,
where $g$ is the string coupling and so all of our D-branes are really
infinitely massive except for the ones that are suddenly massless when
$Z=0$. 

From now we will assume that massless decay products lead to
marginal stability. When we see unexpected results for the moduli
spaces later on this paper it should be noted that things cannot be
improved by changing the rules for the treatment of these massless
D-branes.

Anyway, what remains after the two massless branes $\O_E(0,-1)$ and
$\O_E(-1,0)$ break off is a marginally bound state of $\cE[1]$ and
$\O_E(-1,-1)[2]$. Thus, as in the orbifold case, the 0-brane decays
into a direct sum of fractional branes and this is the only polystable
state.

So, at the exoflop limit point, the ruled surface $E$ has been
replaced by a point. This situation is similar to the orbifold of
section \ref{ss:orb}. Note, in particular that the 0-branes do not see
the exoflop geometry. There is no deformation of the 0-brane on the
collapsed $E$ which could be interpreted as moving into the external
$\P^1$.  The only deformation of the direct sum of the 4 fractional
branes consists of switching $p$ on which corresponds to moving away
from $E$ into the bulk of $X$. This can proven by noting that the
quivers we consider can be interpreted as ``Ext-quivers'', where the
arrows denote Ext$^1$'s between the simple one-dimensional
representations. Any deformation within the category of quiver
representations of a direct sum of simple objects will be given, to
first order, by an extension. Thus we really can only deform a D-brane
by changing the values associated to the arrows.

One explanation for the loss of the exoflopped $\P^1$ is as follows.
In order to see the $\P^1$ sticking out of the \CY\ threefold we would
need to move in the K\"ahler moduli space {\em beyond\/} the stage
where $E$ collapses to a point as in the flop case of section
\ref{s:flop}. Here, $E$ collapses to a point just as we hit the limit
point in the moduli space and there is nowhere further to go. Actually
we may well be able to ``see'' the $\P^1$ using different D-brane
probes as will become apparent in section \ref{s:P1}.

\subsubsection{The finite volume case}

Now consider what happens when we give $X$ finite size. The situation
is very similar to the orbifold of section \ref{ss:orb} in that it now
becomes impossible to shrink $E$ to a point. In particular, let us try
to follow the discriminant locus where $\O_E(-1,0)$ and $\O_E(0,-1)$
are massless as $X$ acquires finite size. Introduce coordinates on the
moduli space
\begin{equation}
z = y^{-1} = \frac{\tilde a\tilde d\tilde e\tilde f}{\tilde p^2\tilde b^2},
    \qquad w = \frac{\tilde a^2\tilde b\tilde c}{\tilde p^4}.
\end{equation}
The coordinate $w$ controls the overall size of the \CY\ threefold $X$
with $w=0$ being the large radius limit. The coordinate $z$ (and thus
$y$) coincides with the same coordinate in the previous subsection
when $w=0$. We want to allow $w$ to become small but nonzero.

The discriminant can be computed as
\begin{equation}
\Delta = 12800000w^2-262144y^2w^3-1024000yw^2+12288y^2w^2+3200yw-192y^2w-16y+y^2.
\end{equation}
We know that two D-branes become massless for solutions of this
equation near $(y,w)=(0,0)$ as this is where the extremal transition
occurs and we know from above that these must be $\O_E(-1,0)$ and
$\O_E(0,-1)$ (using the notation from the previous section).

As $w$ is switched on, the expression for the central charges
(\ref{eq:Zmn}) needs to be reinterpreted a little. The constant 1 is
no longer an exact period and $t_1$ is forced equal to $t_2$. We write
the three independent periods in question as $t^0$, $t$ and $t^2$,
and (\ref{eq:Zmn}) becomes
\begin{equation}
  Z(\O_E(-m,-n)) = (\ff76-m-n+mn)t^0 + (n+m-2)t +t^2, \label{eq:Zmn2}
\end{equation}
To stay on the discriminant we require
\begin{equation}
  Z(\O_E(-1,0)) = \ff16t^0 - t + t^2 =0,
\end{equation}
i.e., set $t^2=t-\ff16t^0$. Given any small $w$ we know that we can fix $y$
such that we stay on the discriminant in this fashion. This is because
the extremal transition can be made to occur for any large but finite volume of
the \CY\ threefold.

Quiver c) in figure \ref{fig:exo1} will
again be the important one. We have
\begin{equation}
\begin{split}
  Z(\O_E(-1,-1)[2]) &= t\\
  Z(\cE[1]) &= t^0-t.
\end{split}
\end{equation}
When $w$ is switched on and kept away from branch cuts, an analysis
very similar to section \ref{s:orb} and \cite{me:min-d} shows that
the component of the K\"ahler form that measures the areas of the
rulings in $E$ becomes strictly positive. The mirror map dictates that
the relevant component of $B+iJ$ is given by the ratio of periods $t/t^0$. That is,
\begin{equation}
  \Im(t/t^0)>0. \label{eq:tim}
\end{equation}
We know that the phases of the central charges $\xi(\O_E(-1,-1)[2])$
and $\xi(\cE[1])$ are equal when $z=w=0$ since that corresponds to a
contracted del Pezzo surface.\footnote{It is believed this a general
  result and can be proven for contracted exceptional sets in
  orbifolds \cite{Asp:theta}. In this case the quadric surface can be
  obtained from a partial resolution of an orbifold, as in
  \cite{MP:AdS}, and thus the result is proven.} It follows that they
are almost equal when $w$ is small and thus (\ref{eq:tim}) gives
\begin{equation}
  \xi(\O_E(-1,-1)[2]) >\xi(\cE[1]).
\end{equation}
This means that $\O_E(-1,-1)[2]$ and $\cE[1]$ can form a stable bound
state $\cG(f)$ from the triangle
\begin{equation}
\xymatrix@1{
  \O_E(-1,-1)[2]\ar[rr]|{[1]}^(0.6)f&&\cE[1]\ar[dl]\\
  &\cG(f)\ar[ul]&
}
\end{equation}
where $f$ comes from the diagonal map in figure \ref{fig:exo1}c). That
is, $f$ is any linear combination of
$u_0v_0,u_0v_1,u_1v_0,v_0v_1$. The moduli space of objects $\cG(f)$ is
therefore $\P^1\times\P^1$, i.e., $E$.

So, to recap, when we lie on the discriminant locus for small values
of $y$ and $w$ and we consider a 0-brane on the quadric surface $E$,
this D-brane becomes marginally stable against a decay into two
massless D-branes and $\cG(f)$. We have a polystable D-brane
\begin{equation}
  \O(0,-1)\oplus \O(-1,0)\oplus \cG(f). \label{eq:polyex}
\end{equation}
But the moduli space of such polystable D-branes is the moduli space
of $\cG(f)$ which is precisely $E$. So our 0-brane probe says that $E$
has {\em not\/} shrunk down to a point.

In one sense this is an expected result. The criterion for when $E$
has shrunk down by measuring areas says that it cannot when $X$ has
finite volume. So the 0-brane probe picture agrees. That said, in
another sense this is a very surprising result. When the two D-branes
become massless as we hit the discriminant we have an extremal
transition to another \CY\ threefold $Z$. From $Z$'s point of view,
the extremal transition occurs because of a singularity induced by a
deformation of complex structure. For a large generic value of the
K\"ahler form, the moduli space of a 0-brane on $Z$ is simply $Z$. So,
from $Z$'s point of view the 0-brane probe says that $E$ has shrunk to
a point at the moment the extremal transition happens. But from $X$'s
point of view it has not. In the extremal transition the volume of $E$
appears to jump discontinuously from some nonzero size to zero.

A resolution of this apparent paradox might come from the fact that we have not
considered the effects of a nonzero string coupling. The idea that
D-branes have a moduli space is a strictly conformal (or topological)
field theory notion for zero string coupling. Once the string coupling
is turned on we need to quantize the D-branes and, if the moduli space
is reasonably well-defined, for the ground states this amounts to
computing the cohomology of the moduli space. The resulting spectrum
of D-branes can certainly jump as we move around the moduli space of
theories. This happens all the time in wall-crossing phenomena such as
\cite{Jafferis:2008uf} and there are no paradoxes here. Matters become
a good deal more complicated when non-ground-states for the D-branes
are considered. The notion of a location of a 0-brane is
contained in the complete string theory. But this is very difficult to
analyze so we will not try to pursue this here.

Let us emphasize that the peculiar behaviour arising from the
impossibility of shrinking $E$ down does not arise from our rules for
treating massless D-branes. The moduli space of the polystable D-brane
(\ref{eq:polyex}) arises from the moduli space of the massive part
$\cG(f)$. Whether we keep the massless parts bound or not will not
collapse the moduli space down to a point.


\section{Probing a Hybrid $\P^1$} \label{s:P1}

\subsection{Basic geometry}

In the previous section we failed to ``see'' the $\P^1$ sticking out
of the \CY\ threefold in an exoflop phase. Recall that this $\P^1$ is
actually a hybrid model in the sense that there is a Landau-Ginzburg
(orbifold) theory over every point in the $\P^1$. It is natural to ask
whether a D-brane probe can ever see the geometry of such a hybrid
phase. That is, can be find a D-brane whose moduli space is
$\P^1$?\footnote{Perhaps if we were more ambitious we could try to
  find a {\em fat\/} $\P^1$ but we will not discuss this here.}

One easy way to obtain a hybrid model is to consider a \CY\ threefold
$X$ that has a fibration $\pi:X\to\P^1$ with generic fibre a K3
surface. Assume for simplicity that this fibration has a section
$C\subset X$. If we vary the K\"ahler form to shrink down $X$ while
keeping $C$ constant we will generically pass into a hybrid model of the
desired type. Each K3 fibre will become a Landau--Ginzburg
theory. Imagine a 4-brane corresponding to $\O_S$ where $S$ is a
generic K3 fibre. The moduli space of such a brane should indeed be
$\P^1$. But this is not a very impressive result --- we have not
really probed anything about the hybrid phase since our result has an
interpretation purely in terms of the geometry of the \CY\ phase.

Instead we could look at an example where the $\P^1$ has no geometric
manifestation in the \CY\ phase. To this end let $X$ be a complete
intersection of two generic cubic equations in $\P^5$.
The toric geometry used to deal with complete intersections is
inherent in Witten's original phases work \cite{W:phase} and was
described further in \cite{Boris:m,AG:gmi}. $\cA$ is a collection of 8 points
in $\R^7$:
\begin{equation}
\begin{array}{c|ccccccc}
p_0&0&1&\pz0&\pz0&\pz0&\pz0&\pz0\\
p_1&1&0&\pz0&\pz0&\pz0&\pz0&\pz0\\
x_0&0&1&\pz1&\pz0&\pz0&\pz0&\pz0\\
x_1&0&1&\pz0&\pz1&\pz0&\pz0&\pz0\\
x_2&0&1&\pz0&\pz0&\pz1&\pz0&\pz0\\
x_3&1&0&\pz0&\pz0&\pz0&\pz1&\pz0\\
x_4&1&0&\pz0&\pz0&\pz0&\pz0&\pz1\\
x_5&1&0&-1&-1&-1&-1&-1
\end{array} \label{eq:p33}
\end{equation}
A triangulation of this point set defines a noncompact \CY\ toric
variety $V_\Sigma$. Define the worldsheet superpotential
\begin{equation}
  W = p_0f_0 + p_1f_1,
\end{equation}
which is an invariant function on $V_\Sigma$, where $f_0$ and $f_1$
are cubic polynomials in $x_0,\ldots,x_5$. Again we define $X_\Sigma$
as the critical point set of $W$:
There are two triangulations of $\cA$ leading to two phases. We label
them by $B_\Sigma$, the Alexander dual of the Stanley-Reisner ideal
$I_\Sigma$ (which is the ideal associated to the variety
removed from affine space prior to quotienting to form the toric variety).
\begin{itemize}
\item The \CY\ phase: $B_\Sigma=(x_0,x_1,x_2,x_3,x_4,x_5)$. $X_\Sigma$
  is a smooth threefold given by the complete intersection of two
  cubics in $\P^5$.  The $\P^5$ has homogeneous coordinates
  $[x_0,\ldots,x_5]$.
\item The hybrid phase: $B_\Sigma=(p_0,p_1)$. $X_\Sigma$ is a (fat) $\P^1$
  with homogeneous coordinates $[p_0,p_1]$. Over each point in this
  $\P^1$ there is a cubic Landau-Ginzburg orbifold theory with fields
  $x_0,\ldots,x_5$.
\end{itemize}

If life were the same as the flop we would have a 0-brane moduli
space of a \CY\ threefold in the \CY\ phase and a 0-brane moduli space
of $\P^1$ in the hybrid phase. This is not at all what
happens. Applying the discussion at the start of section \ref{s:orb},
it is even harder to destabilize the zero brane since we are shrink
down the {\em entire\/} 3-dimensional \CY\ when passing to the hybrid
phase. Na\"\i vely we might expect the angle $\theta_0$ in figure
\ref{fig:flop} to be $2\pi$. This implies that the 0-brane fails to
decay even in the limit of the hybrid. This is the analogue of the
statement that the 0-brane is stable for the Gepner model of the
quintic for which evidence has been given in
\cite{Wal:LGstab,Brunner:2005fv}.

So it would seem that the moduli space of 0-branes is {\em always\/}
the \CY\ threefold irrespective of where we are in the moduli
space. However, from the hybrid's point of view we are only shrinking down
a curve when moving out of its phase and so we would expect life to be
similar to the flop. That is, the geometry-probing D-brane should be
stable only in the hybrid phase. We should therefore expect {\em
  two\/} different classes of D-branes --- the 0-brane which always
gives the \CY\ picture and another probe that gives the $\P^1$ picture
only in the hybrid phase. This is exactly what we find below.

\subsection{Equivalences between D-branes between the 2 phases}

The mathematical machinery required to understand D-branes for
complete intersections in toric varieties is based on matrix
factorizations \cite{KL:Mfac} and was discussed in
\cite{Orlov:mfc,me:csalg,HHP:linphase}. The first question we wish to
address is how to identify D-branes in one phase with D-branes in the
other. In particular, we need the large \CY\ interpretation of a
certain D-brane in the hybrid phase.

We begin with a tilting object for $V_\Sigma$. Let $S$ be the
homogeneous coordinate ring $k[p_0,p_1,x_0,\ldots,x_5]$. Let
\begin{equation}
  T = S\oplus S(-1)\oplus\ldots\oplus S(-5).
\end{equation}
This is a tilting object for $V_\Sigma$ in both phases.

Now add a second grading to the picture which is associated to the
$\GU(1)$ $R$-charge of the gauged linear sigma model. Let $p_0$ and
$p_1$ have bigrading $(2,-3)$ and $x_0,\ldots,x_5$ have bigrading
$(0,1)$. So $W$ has grade $(2,0)$ as one would expect for the
worldsheet superpotential.

Following \cite{Orlov:mfc,me:csalg} we define the category of matrix
factorizations of $W$:
\begin{equation}
  \textrm{DGr}S(W) = \left.\frac{\DC(\textrm{gr-}S')}{\mf{Perf}(S')}
              \right|_{[1]=(1,0).}  \label{eq:DGrS}
\end{equation}
where $S'=S/(W)$ and
$\mf{Perf}(S')$ is the triangulated subcategory of
$\DC(\textrm{gr-}S')$ given by objects with a {\em finite\/} free
resolution. We refer to \cite{Orlov:mfc} for a careful treatment of
the boundedness of the various derived categories involved.

The subscript in (\ref{eq:DGrS}) refers to the fact that we must
identify the homological grading with the $R$-charge. That is, in the
triangulated category $\textrm{DGr}S(W)$ we identify the translation
functor $[1]$ with the grading shift $(1,0)$. Because of this
identification will we rewrite $S(a,b)$ as $S(b)[a]$ when we give
explicit matrix factorizations.

In practice, the mapping between $S'$-modules and matrix
factorizations given by the equivalence (\ref{eq:DGrS}) is quite easy
to obtain. Given an $S'$-module $M$ compute a minimal free resolution.
This resolution will be infinite but asymptotically cyclic of period 2
as discovered by Eisenbud \cite{Eis:mf}. The matrix factorization can
be read off from this asymptotic form.

The category of D-branes in the phase associated to a triangulation
$\Sigma$ is then the quotient of $\textrm{DGr}S(W)$ by
$T^\Delta_\Sigma$, where $T^\Delta_\Sigma$ is the sub-triangulated category of
matrix factorizations generated by modules annihilated by powers of
$B_\Sigma$ \cite{HHP:linphase}. Of course, it is a basic property of
topological B-type D-branes that they should not depend on $\Sigma$.

\def\sOX{\mathsf{O}_X}
There is an equivalence of categories \cite{Eis:mf,Gull:Ext,AG:McExt,me:csalg}
\begin{equation}
  \textrm{DGr}S(W) \cong \DC(\textrm{gr-}A),  \label{eq:StoA1}
\end{equation}
where 
\begin{equation}
  A = \frac{k[x_0,x_1,\ldots,x_5]}{(f_0,f_1)}.
\end{equation}
This equivalence leads to the identification of
$\textrm{DGr}S(W)/T_\Sigma^\Delta$ with the derived category of
coherent sheaves on the complete intersection $X$ in the \CY\ phase.

Let $\O_X$ be the structure sheaf of our \CY\ threefold.  This
corresponds to the $A$-module $A$ itself in $\DC(\textrm{gr-}A)$.
Following the algorithm of \cite{AG:McExt}, $\O_X$ is identified as
the cokernel of the map $S(-3)^{\oplus2}\to S$ given by
$(f_0,f_1)$. Thus the sheaf $\O_X$ is associated with the $S$-module
$\sOX$ defined as the cokernel of the map.
\begin{equation}
\xymatrix@1@C=15mm{
  S(-3)^{\oplus 6}\ar[r]^-{f_0,f_1}& S}
\end{equation}
$\sOX$ is an $S'$-module since it is annihilated by $W$.  If we
compute a free resolution of $\sOX$ {\em as an $S'$-module\/} we
obtain a chain complex that is infinite in length but that can be
repackaged as the matrix factorization:
\begin{equation}
 \xymatrix@R=0mm@C=20mm{
   S(-3)&S(0)\\
   \oplus\ar@<1mm>[r]^{\left(\begin{smallmatrix}f_0&f_1
             \\-p_1&p_0\end{smallmatrix}\right)}
&\oplus\ar@<1mm>[l]^{\left(\begin{smallmatrix}p_0&-f_1
             \\p_1&f_0\end{smallmatrix}\right)}\\
   S(-3)&S(-6)[2]}  \label{eq:Oxmf}
\end{equation}

\def\mw{\mathsf{w}}
The other key object we define is $\mw$ which is given by the
cokernel of the map
\begin{equation}
\xymatrix@1@C=20mm{
  S(-1)^{\oplus 6}\ar[r]^-{x_0,x_1,\ldots,x_5}& S}.
\end{equation}
In the hybrid phase $\mathsf{w}$ represents the structure sheaf of the
zero section of the bundle of $\P^1$. That is, $\mathsf{w}$ represents
the 2-brane wrapping $\P^1$.

A free resolution of the $S'$-module $\mathsf{w}$ is infinite in
length but can be repackaged as a matrix factorization with matrices
of dimension 32:
\begin{equation}
  \xymatrix@R=0mm@C=20mm{
    &S\\
    S(-1)^{\oplus6} &\oplus \\
    \oplus & S(-2)[2]^{\oplus15}\\
    S(-3)[2]^{\oplus20} \ar@<2mm>[r] & *+(20,0){\oplus}\ar@<2mm>[l]\\
    \oplus & S(-4)[4]^{\oplus15}\\
    S(-5)[4]^{\oplus6} & \oplus\\
           & S(-6)[6]}  \label{eq:32m}
\end{equation}

Now, suppose we are in the large radius \CY\ phase. We can represent
the 6-brane that wraps $X$ by $\O_X$ and thus the matrix factorization
(\ref{eq:Oxmf}). The object $\mathsf{w}$ is annihilated by $B_\Sigma$
and thus represents a trivial D-brane. In terms of representing the
6-brane by a matrix factorization we are free to add on (with mapping
cones) arbitrary collections of $\mathsf{w}$'s (and its shifts in
gradings) to (\ref{eq:Oxmf}). The idea of \cite{HHP:linphase} is that
there should be one preferred presentation of $\O_X$ as an $S$-module
that can be naturally transported into the other phase so we can
identify it with matrix factorizations there. This is equivalent
\cite{bergh04:nc,me:toricD} to stating that it is resolved only in
terms of summands of our tilting object. That is, we only want free
$S$-modules of the form $S(n)$, where $-5\leq n\leq0$. In
particular, the $S$-module $\sOX$ is {\em not\/} the preferred
presentation of $\O_X$ as it contains an $S(-6)$.

The trick is to find a map between $\O_X$ and some grade-shifted
$\mathsf{w}$ so that the $S(-6)$ terms cancel when we construct the
mapping cone. We may introduce
polynomials $h_i$ and $g_{ij}$ such that
\begin{equation}
{W}=\sum_{i=0}^5 h_ix_i, \quad f_j=\sum_{i=0}^5g_{ij}x_i.
\end{equation}
Then consider extending the following map between resolutions:
\begin{equation}
\xymatrix@C=15mm{
\ldots\ar[r]&S(-5)^{\oplus6}\ar[r]^{h_0,\ldots,h_5}
  \ar[d]^{(g_{ji})}&S(-6)[2]\ar[d]^1\ar[r]&\coker(h_0,\ldots,h_1)\\
\ldots\ar[r]&S(-3)^{\oplus2}\ar[r]^{p_0,p_1}&S(-6)[2]\ar[r]&\coker(p_0,p_1)
}
\end{equation}
This leads to a map between the matrix factorizations above equivalent
to a map $\mw[-4]\to\O_X$ which gives the necessary cancellation
between the degree $-6$ pieces. That is, we 
have a triangle
\begin{equation}
\xymatrix@C=5mm{
  \mw[-4]\ar[rr]&&\sOX\ar[dl]\\
  &\mathsf{Q}_X\ar[ul]|{[1]}} 
\end{equation}
or, equivalently,
\begin{equation}
\xymatrix@C=5mm{
  \mw[-3]\ar[rr]|{[1]}&&\sOX\ar[dl]\\
  &\mathsf{Q}_X\ar[ul]}  \label{eq:triO}
\end{equation}
where $\mathsf{Q}_X$ represents the 6-brane $\O_X$ in a form that can be
carried into the hybrid phase.

Let us pass to the hybrid phase. Now it is $\sOX$ which is annihilated
by $B_\Sigma$ and so $\sOX$ is now trivial. In the category of
D-branes in the hybrid phase the triangle (\ref{eq:triO}) represents
an isomorphism $\mathsf{Q}_X\cong\mw[-3]$. We have therefore proven
\begin{prop}
  Under the equivalence between D-branes in \CY\ and hybrid phases,
  the 6-brane corresponding to the structure sheaf $\O_X$ is
  mapped\footnote{This mapping is not unique. The mapping we give here
    is the one dictated by the particular tilting object we use.} to
  the matrix factorization $\mw[-3]$, where $\mw$ is the $32\times32$
  matrix factorization given by (\ref{eq:32m}) and represents the
  2-brane wrapping the $\P^1$ zero-section of the Landau--Ginzburg
  fibration over $\P^1$.
\end{prop}
Let us denote this correspondence $\O_X\leftrightsquigarrow\mw[-3]$.

We could also consider a morphism
\begin{equation}
\xymatrix@C=15mm{
\ldots\ar[r]&S(-1)^{\oplus6}\ar[r]^{x_0,\ldots,x_5}
  &S(0)\ar[r]&\mw\\
\ldots\ar[r]&S(-3)^{\oplus2}\ar[u]^{(g_{ij})}
  \ar[r]^{f_0,f_1}&S(0)\ar[u]^1\ar[r]&\sOX
}  \label{eq:mor2}
\end{equation}
which would cancel out the $S(0)$'s in the mapping cone. Twisting
everything by $S(1)$ this gives a triangle
\begin{equation}
\xymatrix@C=5mm{
 \sOX(1)\ar[rr]&&\mw(1)\ar[dl]|{[1]}\\
  &\mathsf{R}_X\ar[ul]}
\end{equation}
where $\mathsf{R}_X$ is an $S$-module that represents the sheaf
$\O_X(1)$ in the \CY\ phase in a form suitable for transport into the
hybrid phase. In the hybrid phase, $\sOX(1)$ becomes trivial and
therefore we have a correspondence
\begin{equation}
  \O_X(1)\leftrightsquigarrow\mw(1)[-1].
\end{equation}

Now consider what, in the \CY\ phase would correspond to $\mw(2)$ in
the hybrid phase. Looking at (\ref{eq:mor2}) twisted by $S(2)$ we see
that we need to get rid of $S(1)^{\oplus6}$ and $S(2)$ to yield a resolution
purely in terms of the tilting summands. We may remove the $S(2)$'s by
coning with $\sOX(2)$ and then the six $S(1)$'s by coning with
$\sOX(1)$'s. When we pass to the \CY\ phase the $\mw(2)$ becomes trivial
and we are left with a correspondence
\begin{equation}
  \mw(2) \leftrightsquigarrow \left(
\xymatrix@1{\O_X(1)^{\oplus6}\ar[r]&\poso{\O_X(2)}} \right)[1],
\end{equation}
where the dotted line denotes position 0 in the complex.
Let $\Omega^n$ denote the $n^{\textrm{th}}$ exterior power of the
cotangent sheaf of $\P^5$ and let $\Omega^n_X$ denote
$\Omega^n\otimes\O_X$. Then the above becomes $\mw(2)
\leftrightsquigarrow \Omega^1_X(2)[2]$. Continuing in this fashion we
get a correspondence
\begin{equation}
  \mw(n) \leftrightsquigarrow \Omega^{n-1}_X(n)[n],\qquad\hbox{for $n=1,2,3$.}
         \label{eq:mwn}
\end{equation}
Note that (\ref{eq:mwn}) fails for $n=4$ because maps containing $p_i$
in (\ref{eq:32m}) start to make the resolution of $\mw$ in
(\ref{eq:mor2}) differ from the Koszul resolution.\footnote{This fact
  ruins any hope of finite monodromy around this limit that would have
  been seen in, for example, the Landau-Ginzburg phase of the
  quintic.}

\subsection{D-branes with moduli space $\P^1$}

Consider a $\P^1$ with homogeneous coordinate ring $S=k[x_0,x_1]$ and
let $p$ be a point in this curve with homogeneous coordinates
$[u_0,u_1]$. If $\O$ is the structure sheaf of the curve and $\O_p$ is
the skyscraper sheaf of $p$ we have a short exact sequence
\begin{equation}
\xymatrix@1{
  0\ar[r]&\O(-1)\ar[rr]^-{u_0x_0+u_1x_1}&&\O\ar[r]&\O_p\ar[r]&0.} 
          \label{eq:Op1}
\end{equation}
The variables $p_0$ and $p_1$ in the homogeneous coordinate ring serve
as the homogeneous coordinates of $\P^1$ in the hybrid phase. We
already know that $\mw$ represents the structure sheaf of this
$\P^1$. Let $p$ be a point on this $\P^1$ with homogeneous coordinates
$[u_0,u_1]$ and let $\cP_p$ be the cokernel of a linear combination of
these coordinates corresponding to the point $p$. Given
that $p$ has charge $(2,-3)$, we have a triangle
\begin{equation}
\xymatrix{
&\cP_p\ar[dl]|-{[1]}&\\
\mw(3)[-2]\ar[rr]^-{u_0p_0+u_1p_1}&&\mw\ar[ul]} \label{eq:p1a}
\end{equation}
as the analogue of (\ref{eq:Op1}). Thus $\cP_p$ is the natural
candidate for a brane to probe the geometry of the $\P^1$.

From the previous section we know we may rephrase (\ref{eq:p1a}) in terms of
$\DC(X)$ as
\begin{equation}
\xymatrix{
&\cP_p\ar[dl]|-{[1]}&\\
\Omega_X^2(3)[1]\ar[rr]^-{u_0p_0+u_1p_1}&&\O_X[3]\ar[ul]} \label{eq:p1p}
\end{equation}

It is important to note that it would be misleading to
call $\cP_p$ a 0-brane. Looking at (\ref{eq:p1p}) it is not hard to see
that the K-theory class of $\cP_p$ is not equal to that of a point. We
need a different D-brane to ``see'' the $\P^1$. In all probability,
the true 0-brane probes remain stable everywhere in the moduli space (so
long as we don't cross branch cuts) and so the geometry by this
criterion is always $X$. 

In order for $\cP_p$ to see the geometry of the $\P^1$, it must be
stable. At large radius limit, the phase of a coherent sheaf is given
by
\begin{equation}
  \xi(\cE) = -\ff12\dim\supp(\cE).
\end{equation}
It follows that
\begin{equation}
  \xi(\O_X[3]) - \xi\left(\Omega_X^2(3)[1]\right) = 2,
\end{equation}
and so (assuming these sheaves are stable) $\cP_p$ is {\em unstable}
from (\ref{eq:p1p}).

So, at the large radius limit, evidence of the $\P^1$ of the hybrid
phase does not exist. Naturally we would like to know if it appears as
we move towards the hybrid phase. Lacking a practical rigorous
stability criterion we will just consider the triangle (\ref{eq:p1p})
to determine when $\cP_p$ is stable. 

We need to determine the exact values of the phases $\xi$ as we move
in the moduli space of $B+iJ$. This is very similar to the analysis of
the quintic done in \cite{AD:Dstab}. Since $h^{1,1}(X)=1$, let
$B+iJ=tx$, where $x$ is dual to the divisor class $x_0=0$ in $X$. 

Introduce the algebraic coordinate 
\begin{equation}
  z = \frac{\tilde a_1\tilde a_2\tilde a_3\tilde a_4\tilde a_5\tilde
    a_6}{\tilde p_0^3\tilde p_1^3}.
\end{equation}
The periods then solve the Picard--Fuchs equation $\square\Phi=0$, where
\begin{equation}
  \square = \left(z\frac{\partial}{\partial z}\right)^4 -
      27^2z\left(z\frac{\partial}{\partial z}+\frac13\right)^2
      \left(z\frac{\partial}{\partial z}+\frac23\right)^2.
\end{equation}
It is useful to put $z = (9\psi)^{-3}$, where $\psi=0$ in the
limit of the hybrid phase. 

The central charges of the line bundles $\O(n)$ are then given by
\begin{equation}
\begin{split}
Z(\O(n)) &= \int_X e^{-tx} e^{nx}(1+\ff14x^2)+\ldots\\
  &= -\ff34(t-n)(2t^2-4tn+3+2n^2)+\ldots,
\end{split}
\end{equation}
which allows them to be identified in periods as was done in
\cite{AD:Dstab,me:TASI-D}.

\begin{figure}
\begin{center}
\includegraphics[width=140mm]{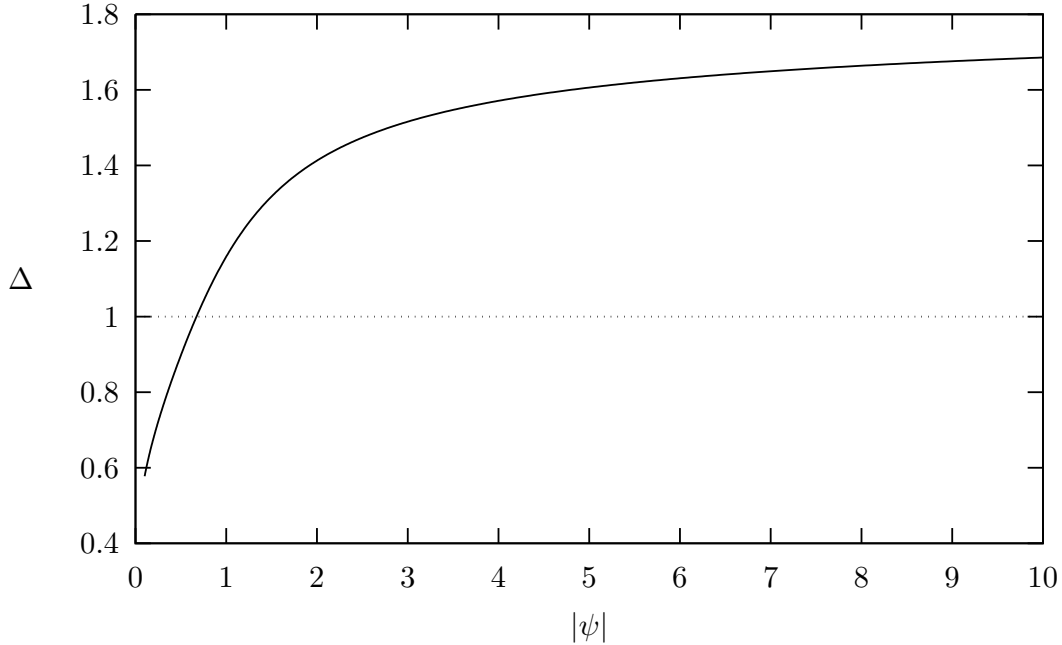}
\end{center}
\caption{Difference in phases for decay of $\cP_p$ on $\P^1$.} \label{fig:graph}
\end{figure}

A lengthy numerical computation then yields the difference in the
phases $\Delta = \xi(\O_X[3]) - \xi\left(\Omega_X^2(3)[1]\right)$ as a
function of $\psi$. In figure \ref{fig:graph} we plot how $\Delta$
varies with $|\psi|$ where we set $\arg(\psi)=-i\pi/3$ to keep away
from branch cuts. One sees that for sufficiently small $|\psi|$,
$\Delta<1$ and so $\cP_p$ is stable relative to the triangle
(\ref{eq:p1p}). 

Of course to be rigorous we need to consider {\em all\/} possible
triangles. We also need to be sure that both $\O_X$ and
$\Omega_X^2(3)$ are stable when we apply our criterion for
stability. We cannot address either of these issues rigorously but one
should note that both $\O_X$ and $\Omega_X^2(3)$ are ``fractional
branes'' that become massless and induce singularities in the
conformal field theory and some points in the moduli space. We
therefore know that they are stable ``nearby'' in some sense. Anyway,
we will assume that the only triangle needed for checking stability is 
(\ref{eq:p1p}).

Assume first $\Delta<1$ in figure \ref{fig:graph}. Then $u_0=u_1=0$,
corresponding to $\cP_p=\Omega_X^2(3)[2]\oplus\O_X[3]$, is unstable,
and we have an obvious $\C^*$-action, $(u_0,u_1)\to(\lambda
u_0,\lambda u_1)$, yielding isomorphism classes of $\cP_p$. The moduli
space of $\cP_p$ is thus $\P^1$. If $\Delta=1$ then
$\cP_p=\Omega_X^2(3)[2]\oplus\O_X[3]$ is polystable and no nonzero
values of $u_0$ or $u_1$ yield polystable objects so the moduli space
of $\cP_p$ is a point. Thus at $\Delta=1$ the $\P^1$ shrinks to a
point and for $\Delta>1$ we have no $\P^1$ at all.

So we do seem to have a D-brane probe whose moduli space is the
$\P^1$ base of the hybrid.  It would appear then, that in the hybrid
phase both the original 0-brane $\O_p$ and this new D-brane probe
$\cP_p$ are stable. One of them, $\O_p$, has a moduli space given by
the \CY\ threefold while the other, $\cP_p$ has a moduli space of
$\P^1$.


\section{Discussion}  \label{s:conc}

In sections \ref{s:flop}, \ref{s:orb} and \ref{s:P1} we considered
shrinking, respectively, a curve, a divisor and the whole
threefold. As one might expect from the na\"\i ve angle argument from
the beginning of section \ref{s:orb}, we see roughly that 0-brane decay
happens as we pass between phases, as we hit the limit of a phase or
never, respectively.

In addition, in the hybrid case we could find a D-brane whose
moduli space was $\P^1$ so long as we remained in the hybrid
phase. This is consistent with the point of view that a
shrinking $\P^1$ supports no stable ``0-branes'' as we pass beyond
the phase boundary. This suggests a nice resolution of the fact that
this model is both 3-dimensional and 1-dimensional. 

In the exoflop case we saw that the 0-brane probe never sees the
external $\P^1$. Presumably, from the results of section \ref{s:P1},
there is another D-brane which does probe this $\P^1$ correctly so
long as we are in the exoflop phase. That is, just like the hybrid
case, we need two different D-brane charges to fully probe the
geometry. It would be interesting to find this expected extra D-brane.

Aside from this expected result there are some surprising results
coming from finite volume effects. We recover, from a brane probe
point of view, the minimum distance idea of \cite{me:min-d} that
exceptional divisors can only be shrunk down to zero size if the
volume of the \CY\ threefold is infinitely large.  This turns out to
be quite a surprise in the extremal transition case since this results
in an even more discontinuous change in geometry than expected for a
topology transition. It would be very interesting to see exactly what
happens to this discontinuity when the string coupling is turned on.


\section*{Acknowledgments}

I wish to thank S.~Guerra, I.~Melnikov, R.~Plesser and A.~Roy for useful
discussions. The author is supported by an NSF grant DMS--0606578.


\end{document}